\documentclass{aa}

\usepackage{natbib}
\usepackage{graphicx}
\usepackage{txfonts}
\usepackage{array}
\usepackage{subfigure} 
\usepackage[skip=5pt]{caption} 
\usepackage[version=3]{mhchem}

\begin{document}

\title{Chemical complexity induced by efficient ice evaporation in the Barnard 5 molecular cloud}

\author{V. Taquet\inst{1} \and E. S. Wirstr{\"o}m\inst{2} \and S. B. Charnley\inst{3} \and A. Faure\inst{4,5}
  \and A. L\'opez-Sepulcre\inst{6,7} \and C. M. Persson\inst{2}}

\offprints{V. Taquet: taquet@strw.leidenuniv.nl}

\institute{Leiden Observatory, Leiden University, P.O. Box 9513,
  2300-RA Leiden, The Netherlands \and
Department of Earth and Space Sciences, Chalmers University of
Technology, Onsala Space Observatory, 439 92 Onsala, Sweden \and
Astrochemistry Laboratory, Mailstop 691, NASA Goddard Space
Flight Center, 8800 Greenbelt Road, Greenbelt, MD 20770, USA \and
Univ. Grenoble Alpes, IPAG, F-38000 Grenoble, France \and
CNRS, IPAG, F-38000 Grenoble, France \and
Department of Physics, The University of Tokyo, 7-3-1 Hongo,
Bunkyo-ku, Tokyo, 113-0033, Japan \and
Institut de Radioastronomie Millimétrique, Grenoble, France
}

\date{Received / Accepted}

\abstract
{
Cold gas-phase water has recently been detected in a cold
dark cloud, Barnard 5 located in the Perseus complex, by targeting
methanol peaks as signposts for ice mantle evaporation. 
Observed morphology and abundances of methanol
and water are consistent with a transient non-thermal evaporation
process only affecting the outermost ice mantle layers, possibly
triggering a more complex chemistry. 
We present the detection of the Complex Organic Molecules (COMs)
{ acetaldehyde (CH$_3$CHO) and methyl formate
(CH$_3$OCHO) as well as formic acid (HCOOH) and ketene (CH$_2$CO),} and
the tentative detection of di-methyl ether (CH$_3$OCH$_3$) towards
{ the methanol hotspot} of Barnard 5 located between two 
dense cores { using} the single dish OSO 20m, IRAM 30m, and
NRO 45m telescopes. 
The high energy {\it cis-} conformer of formic acid is detected,
suggesting that formic acid is mostly formed at the 
surface of interstellar grains and then evaporated. 
The detection of multiple transitions for each species allows us to
constrain their abundances through LTE and non-LTE methods. 
All the considered COMs show similar abundances between $\sim 1$
and $\sim 10$ \% relative to methanol depending on the assumed excitation
temperature. 
The non-detection of glycolaldehyde, an isomer of methyl formate, with
a [glycolaldehyde]/[methyl formate] abundance ratio lower than 6 \%,
favours gas phase formation pathways triggered by methanol evaporation.
According to their excitation temperatures derived in massive
hot cores, formic acid, ketene, and acetaldehyde have been designated as
``lukewarm'' COMs whereas methyl formate and di-methyl ether were {
  defined as} ``warm'' species. 
Comparison with previous observations { of other types of sources}
confirms that ``lukewarm'' and ``warm'' COMs  show similar abundances
in low-density cold gas whereas the ``warm'' COMs tend to be more
abundant than the ``lukewarm'' species in warm protostellar cores. 
This abundance evolution suggests either that ``warm'' COMs are indeed
mostly formed in protostellar environments and/or that ``lukewarm'' COMs
are efficiently depleted by increased hydrogenation efficiency around
protostars. }


\maketitle

\titlerunning{Chemical complexity in B5}
\authorrunning{V. Taquet et al.}

\section{Introduction}

It is known for more than a decade that the early stages of low-mass
star formation are rich in interstellar complex organic molecules
(COMs, i.e. molecules based on carbon chemistry and with { six atoms
  or more}; Herbst \& van Dishoeck 2009) since the detection of
several saturated COMs by \citet{Cazaux2003} and \citet{Bottinelli2004a,
  Bottinelli2007} towards nearby bright Class 0 protostars with
sub-millimeter single-dish telescopes. 
Subsequent interferometric observations of Class 0 protostars confirmed
that the emission of COMs mostly originates from the inner
warm regions of protostellar envelopes, the so-called hot corinos,
the low-mass counterparts of the massive hot cores
\citep{Bottinelli2004b, Kuan2004, Jorgensen2005, Jorgensen2011,
  Maury2014, Taquet2015}. 

The current scenario explaining the detection of { warm ($T > 100$ K)} COMs surrounding
protostars is mostly based on grain surface chemistry followed by
``hot core'' gas phase chemistry. 
In this paradigm, cold ($T \sim 10$ K) ices, containing the first ``parent''
organic molecules like formaldehyde H$_2$CO and methanol CH$_3$OH and
eventually several other COMs (e.g. ethanol C$_2$H$_5$OH, or ethylene
glycol (CH$_2$OH)$_2$), are formed at the surface of interstellar
grains in dark clouds by atom additions on cold dust. 
It has been hypothesized that more complex molecules could also be
formed in { lukewarm} (30 K $< T <$ 80 K) ices through the 
recombination of radicals produced by the UV photolysis of the main
ice components during the warm-up phase in protostellar envelopes.
All the ice content is then evaporated into the gas phase when the
temperature exceeds 100 K \citep{Garrod2006, Garrod2008}. 
Ion-neutral chemistry triggered by the
evaporation of ices, in which ammonia plays a key role, could also be
important for the formation of abundant COMs, such as methyl formate
CH$_3$OCHO or di-methyl ether CH$_3$OCH$_3$ \citep{Taquet2016}. 

The detection of cold methanol, ketene CH$_2$CO, acetaldehyde
CH$_3$CHO, or formic acid HCOOH observed in dark clouds for several
decades \citep{Matthews1985,   Friberg1988, Irvine1989, Irvine1990,
  Ohishi1998} can be explained by the cold surface chemistry scenario mentioned
previously followed by non-thermal evaporation processes. 
However, the UV-induced scenario of COM formation on { lukewarm} ($T \geq
30$ K) interstellar grains, invoked to produce methyl formate and
di-methyl ether, cannot explain the recent first clear detection by \citet{Bacmann2012} of
these two species in a cold ($T \sim 10$ K) prestellar
core, L1689B, shielded from strong UV radiation.  
Several COMs have also been detected by \citet{Oberg2010} and
\citet{Cernicharo2012} towards the core B1-b. However, more
recent interferometric observations suggest that the molecular
emission should rather come from nearby protostars, hydrostatic core
candidates and outflows \citep[see][]{Gerin2015}.  
More recently, \citet{Vastel2014} and \citet{JimenezSerra2016} also
detected CH$_3$OH, HCOOH, CH$_2$CO, CH$_3$CHO, CH$_3$OCHO, and CH$_3$OCH$_3$
towards the prestellar core prototype L1544. A non-LTE 
analysis of the methanol emission towards L1544 suggests that the
emission from COMs would mostly originate from the external part of
the core, with an intermediate density $n_{\rm H} = 4 \times 10^4$
cm$^{-3}$ \citep{Vastel2014}. 
However, subsequent observations of methanol towards 8 other
prestellar cores by \citet{Bacmann2016} suggest { that the cold methanol
emission could be associated with
  denser ($n_{\rm H} \geq 10^5$ cm$^{-3}$) gas}.

New chemical pathways have been proposed to explain the
detection of COMs towards these cold regions. 
\citet{Vasyunin2013} and \citet{Balucani2015} introduced neutral-neutral
gas phase reactions triggered by the non-thermal evaporation of
methanol assumed to be mostly formed at the surface of grains whilst
\citet{Reboussin2014} investigated the effect of heating by
cosmic-rays on the surface formation of COMs. 
\citet{Fedoseev2015} and \citet{Chuang2016} experimentally showed that
surface hydrogenation of CO at low ($T = 15$ K) temperatures leads to
the detection of several COMs through recombination of radicals whose
production is triggered by abstraction reactions of
formaldehyde and methanol. 
However, these experiments tend to produce more glycolaldehyde and
ethylene glycol, not yet detected towards dense clouds, than
methyl formate whilst di-methyl ether cannot be efficiently produced
in their experimental setup.  

COMs have been detected towards only a few dark cloud regions so
far. In this work, we aim at investigating the level of chemical
complexity in the so-called "methanol hotspot" region of the Barnard 5
molecular cloud, one of the two molecular clouds with L1544 where cold
water vapour has been detected with {\it Herschel}
\citep{Caselli2010, Caselli2012, Wirstrom2014}, by observing mm-emission from COMs with 
single-dish telescopes. 
This region, offset from any infrared sources, shows a high abundance
of methanol and water attributed to transient evaporation processes
\citep[see][]{Wirstrom2014}. 
Section 2 describes our observational campaign focusing on detecting
oxygen-bearing COMs. Section 3 presents the spectra and the analysis
to derive column densities and abundances. Section 4 discusses the
results.

\section{Observations}

\subsection{The ``methanol hotspot'' in Barnard 5}

\begin{figure*}
\centering
\includegraphics[width=80mm]{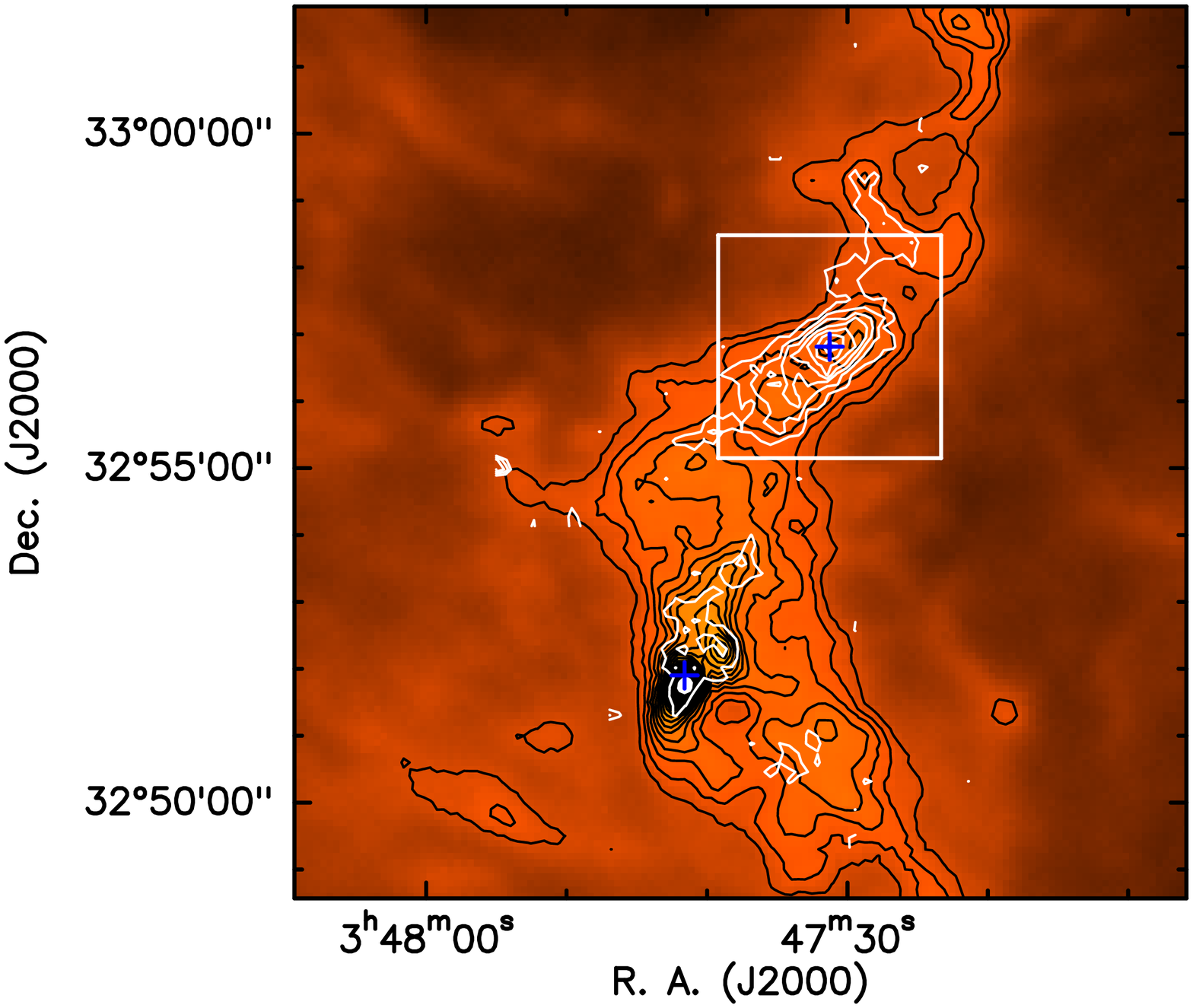}
\includegraphics[width=80mm]{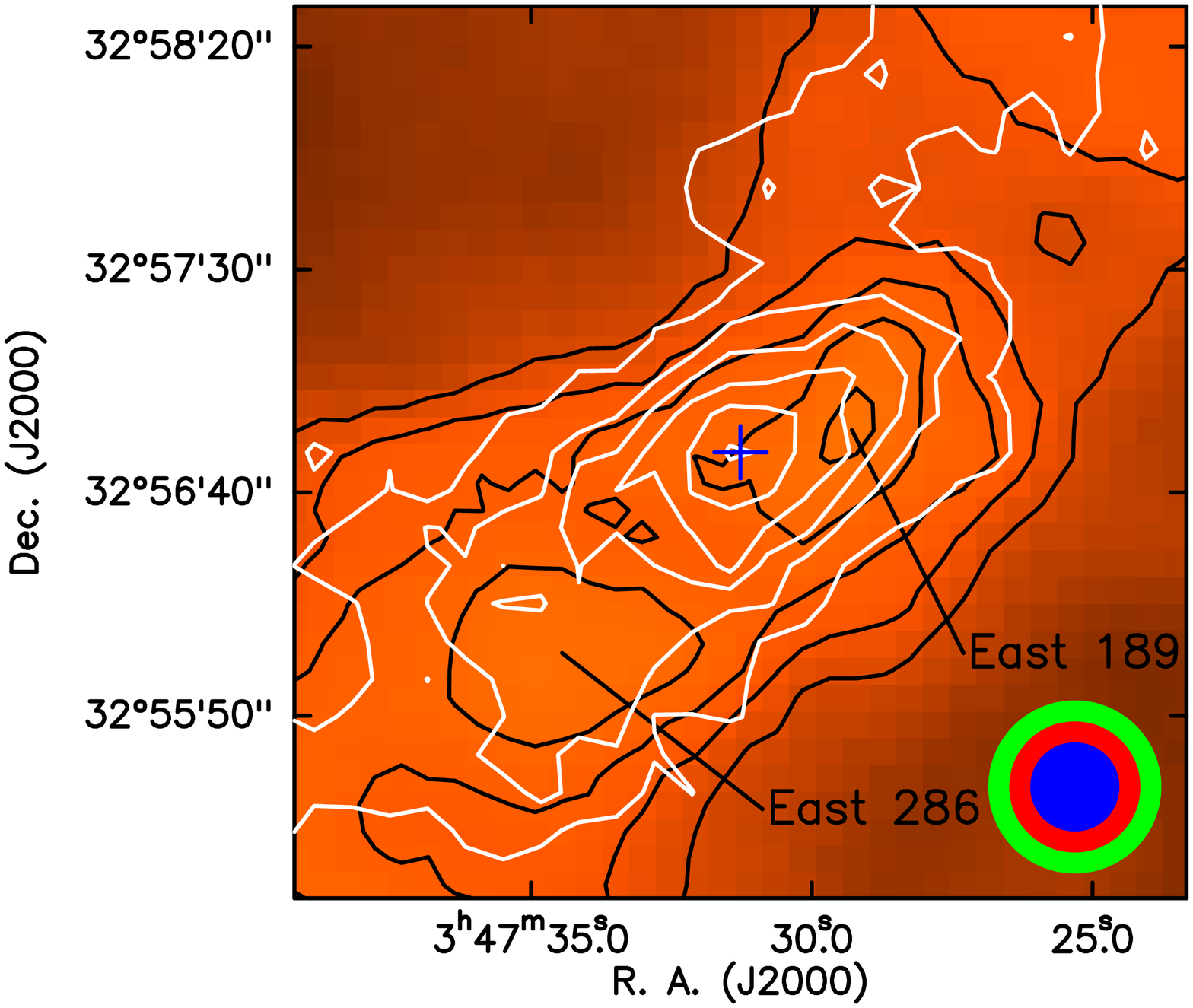}
\caption{ {\it Herschel}/SPIRE map of the Barnard 5 dark cloud at 250
  $\mu$m (orange scale and black contours) and IRAM 30m map integrated
  intensity map of the A$^+$-CH$_3$OH $3_0-2_0$ transition at
  145.103 GHz (white contours in step of 5$\sigma$, { $\sigma$ is
    equal to 50 mK km s$^{-1}$}). The blue crosses
  depict the positions of the Class I protostar IRS1 and the ``methanol hotspot'' of B5. 
The green, red, and blue circles at the bottom right of the right map represent the size of
 the beams of the OSO 20m, the IRAM 30m, and the NRO 45m
 telescopes at 3 mm, respectively. 
The two dense cores surrounding the ``methanol hotspot'' have the following
  properties: East189: $M = 0.5$ $M_{\odot}$, $T_d = 12$ K, $R = 3.6
  \times 10^{-2}$ pc, $n_{\rm{H}} = 3 \times 10^4$ cm$^{-3}$, 
East286: $M = 0.7$ $M_{\odot}$, $T_d = 9.9$ K, $R = 2.7 \times 10^{-2}$
pc, $n_{\rm H} = 1.2 \times 10^5$ cm$^{-3}$.}
\label{map_B5}
\end{figure*}

The Barnard 5 (B5) dark cloud is located at the North-East of the Perseus
complex \citep[235 pc;][]{Hirota2008}. It contains four known protostars of which the most
prominent is the  Class I protostar IRS1. SCUBA continuum emission
maps at 850 $\mu$m carried out with the JCMT only revealed dust emission towards
known protostars and at about 2-4 arcmin north from IRS 1
\citep{Hatchell2005} whilst molecular maps of CO, NH$_3$, and other
species show several chemically differentiated 
clumps (S. B. Charnley in preparation). Methanol emission in B5 displays a particularly
interesting distribution since it shows a bright so-called ``methanol
hotspot''  at about 250 arcsec North-West from IRS 1, a region showing no
infrared sources and no detected sub-mm continuum emission \citep[see
Fig. 1 in][]{Hatchell2005, Wirstrom2014}. 
The subsequent detection of abundant water, with absolute abundances of about
$10^{-8}$, with the {\it Herschel Space Observatory } by
\citet{Wirstrom2014} suggests that efficient non-thermal processes
triggering the evaporation of icy methanol and water are at work. 
The ``methanol hotspot'' therefore represents an ideal target to
detect cold COMs in dark clouds since they are supposed to be formed
from methanol, either at the surface of interstellar grains, or
directly in the gas phase after the evaporation of
methanol. 
The Perseus molecular cloud was recently mapped with {\it
  Herschel} as part as the {\it Gould Belt Survey}
key program \citep{Andre2010} using the photometers PACS and SPIRE in
five bands between 70 $\mu$m and 500 $\mu$m. 
{
Figure \ref{map_B5} presents the Barnard 5 molecular cloud as seen by
{\it Herschel}/SPIRE at 250 $\mu$m. The {\it Herschel} data has been
retrieved from the {\it Herschel Science Archive}. The data were calibrated and
processed with the {\it Herschel} pipeline and are considered as a
level-3 data product.
Dust emission is compared with the  integrated intensity map of the
A$^+$-CH$_3$OH $3_0-2_0$ transition as observed with the IRAM 30m
telescope (this work). 
}
The ``methanol hotspot'' is located between the two dense
cores East 189 and East 286 (Sadavoy, private communication) revealed
by {\it Herschel} for the first time, { and the methanol emission peaks
  at the edge of core East 189}. 
{ East189 and East286 have the following properties derived through a
fitting of their observed Spectral Energy Distribution: 
$M = 0.5$ $M_{\odot}$, $T_d = 12$ K, $R = 3.6   \times 10^{-2}$ pc,
$n_{\rm{H}} = 3 \times 10^4$ cm$^{-3}$, and $M = 0.7$ $M_{\odot}$, $T_d
= 9.9$ K, $R = 2.7 \times 10^{-2}$ pc, $n_{\rm H} = 1.2 \times 10^5$ cm$^{-3}$
\citep{Sadavoy2013}. }

\subsection{Observational details}

We targeted several COMs in different frequency bands at 3 mm using
the NRO 45m, IRAM 30m, and OSO 20m telescopes { and CH$_3$OH in two
frequency bands at 2 mm using the IRAM 30m towards the
  methanol hotspot at R.A. = 3$^{\rm h}$47$^{\rm m}$32$^{\rm s}$.10,
  decl = +32$^{\rm d}$56$^{\rm m}$43$^{\rm s}$.0. } 
Table \ref{log_obs} summarises the properties of our observational data.

Observations with the NRO 45m telescope were carried out in February
2015. We used the TZ1V and TZ1H receivers, connected to the SAM45 
spectrometer, tuned to a frequency of 94 GHz. 
Eight 250 MHz bands between 86.3 and 90.3 GHz and between 98.3 and
102.3 GHz were covered with a spectral resolution of 60 kHz,
corresponding to a velocity resolution of $\sim$ 0.2 km/s at $\sim$ 90
GHz.  
The observations were performed with the position switching
mode. Pointing was checked every 1-1.5 hour on a nearby bright source
and was found to be accurate within 3\arcsec.
The weather conditions were good, although unstable during one night,
and system temperatures were typically 120-180 K. The size of the
telescope beam at the observing frequency is $\sim$18\arcsec, and the
aperture $\eta_{ap}$ and main beam $\eta_{mb}$ efficiencies are $0.30$ and
$0.40$, respectively.
The 45m data were reduced using the Jnewstar software. The scans were
checked individually, flagged and coadded. A polynomial order of 3 was then
fitted over line-free regions to correct for baseline oscillations. 

{ Deep observations with the IRAM 30m telescope towards the
  methanol hotspot were carried out in January 2015 and March 2015. 
The EMIR E090 receiver was used and
  connected to the FTS backend with a 50 kHz resolution.  
Pointing was checked every 1.5 hour on a nearby bright source
and was found to be accurate within 3\arcsec.
Observations were performed with the frequency switching mode,
resulting in a rms of 3-4 mK per 50 kHz spectral channel. 
The weather conditions were average to good, and gave system temperatures
of 70-80 K at 3 mm. 
A low polynomial order of 3 or 4 was then fitted 
over line-free regions to correct for local baseline oscillations.}

{ On-the-fly observations of CH$_3$OH transitions towards the
  Barnard 5 cloud with the IRAM 30m telescope were carried out in
  February 2017 using the position switching mode with a common center
  at IRS1. 
The EMIR E090 and E150 receivers were used and connected to the FTS backend with a 50 kHz
  resolution. 
Pointing was checked every 1.5 hour on a nearby bright source
and was found to be accurate within 3\arcsec.
Position switching observations resulted in a rms of 30-40 mK per 50
kHz spectral channel.  
The size of the telescope beam is about 25-30\arcsec at 3 mm and about
$\sim 15$\arcsec at 2 mm. 
A low polynomial order of 1 was then fitted over line-free regions to
correct for local baseline oscillations. }

{
The 30m data were reduced using the GILDAS/CLASS software. The scans
were checked individually, coadded, and eventually folded to deconvolve the
spectra from the frequency-switching procedure. 
The 1.8 GHz spectral windows covered by our observations are listed in
Table \ref{log_obs}.  
Pointing was checked every 1.5 hour on a nearby bright source
and was found to be accurate within 3\arcsec.
A more comprehensive analysis of the CH$_3$OH maps in the B5 dark
cloud will be presented in a separate publication. 
}

Observations with the Onsala 20m telescope were carried out in
May 2014 and March 2015. We used the dual-polarisation,
sideband-separating 3 mm receiver system \citep{Belitsky2015} to target
four different frequency settings: 89.58 and 95.95 GHz in 2014, and
101.25 and 101.98 in 2015. In 2014  the spectra were recorded using a
fast Fourier transform spectrometer (FFTS) at channel separation 12.2
kHz, covering 100 MHz bandwidth. Dual beam switching mode was used
with an 11\arcmin ~beam throw, and the system temperature varied
between 145-235~K. In 2015, a new FFTS of improved bandwidth was used
at channel separation 19.1 kHz, covering 625 MHz wide spectral
windows. These observations were performed in the frequency switching
mode with a switching frequency of 5 Hz and a throw of 3
MHz. Conditions were good with system temperatures in the range
160-250 K and pointing and focus were checked toward strong continuum
sources after sunrise and sunset for all observations.  
The OSO20m beam full width at half maximum (FWHM) is about 41\arcsec\
at 89 GHz, 38\arcsec\ at 95 GHz and 36\arcsec\ at 101 GHz, and the
main beam efficiency varies with both frequency and source elevation,
resulting in values 0.42-0.50 for the current observations. 
The 20m data were checked and reduced using the spectral analysis
software XS\footnote{Developed by Per Bergman at Onsala Space
  Observatory, Sweden;
  http://www.chalmers.se/rss/oso-en/observations/data-reduction-software}. Linear
baselines were fitted to line-free regions and subtracted from
individual, dual polarisation spectra before averaging them together,
weighted by rms noise. For the frequency switched spectra, which tend
to exhibit low-intensity standing wave features, an additional
baseline of order 3-5 was fitted locally around each line and
subtracted before further analysis.

\begin{table*}[htp]
\centering
\caption{Properties of the observations carried out in this work.}
\begin{tabular}{l c c c c c c c}
\hline
\hline
Frequency range	&	Telescope	&	Beam size	&	df	&	Rms	&	$F_{\textrm{eff}}$	&	$B_{\textrm{eff}}$	&	Targeted molecules	\\
(GHz)	&		&	(arcsec)	&	(kHz)	&	(mK)	&	(\%)	&	(\%)	&		\\
\hline															
83.550-85.370	&	IRAM 30m	&	29	&	50	&	3-4	&	95	&	81	&	CH$_3$OH, CH$_3$OCHO, CH$_3$OCH$_3$, CH$_3$CHO	\\
86.830-88.650	&	IRAM 30m	&	29	&	50	&	3-4	&	95	&	81	&	HNCO, c-HCOOH, CH$_3$OCHO, CH$_3$OCH$_3$	\\
89.532-89.629	&	OSO 20m	&	41	&	60	&	8	&	-	&	49	&	t-HCOOH	\\
89.190-89.370	&	NRO 45m	&	20	&	61	&	3-5	&	-	&	44	&	CH$_3$OCHO	\\
89.350-89.530	&	NRO 45m	&	20	&	61	&	5-6	&	-	&	44	&	CH$_2$DOH	\\
89.530-89.780	&	NRO 45m	&	20	&	61	&	4-5	&	-	&	44	&	CH$_3$OCH$_3$	\\
90.075-90.325	&	NRO 45m	&	20	&	61	&	5-6	&	-	&	44	&	CH$_3$OCHO	\\
95.901-96.001	&	OSO 20m	&	38	&	64	&	6-7	&	-	&	50	&	E-CH$_3$CHO, A-CH$_3$CHO	\\
{ 96.020-97.840}	&	{ IRAM 30m}	&	{ 25}	&	{ 50}	&	{ 27-30}	&	{ 94}	&	{ 80}	&	{ CH$_3$OH}	\\
98.740-98.990	&	NRO 45m	&	18	&	61	&	4-5	&	-	&	42	&	E-CH$_3$CHO, A-CH$_3$CHO	\\
99.200-99.450	&	NRO 45m	&	18	&	61	&	4-5	&	-	&	42	&	CH$_3$OCH$_3$	\\
99.230-101.050	&	IRAM 30m	&	25	&	50	&	3-4	&	94	&	80	&	CH$_3$OCHO, CH$_3$OCH$_3$	\\
100.942-101.558	&	OSO 20m	&	36	&	68	&	5	&	-	&	42	&	p-CH$_2$CO	\\
100.910-101.160	&	NRO 45m	&	18	&	61	&	4-5	&	-	&	42	&	CH$_2$CO, CH$_3$CHO	\\
101.675-102.290	&	OSO 20m	&	36	&	68	&	6	&	-	&	47	&	o-CH$_2$CO	\\
102.510-104.330	&	IRAM 30m	&	25	&	50	&	3-4	&	94	&	80	&	CH$_3$OCHO	\\
{ 143.800-145.620}	&	{ IRAM 30m}	&	{ 16}	&	{ 50}	&	{ 35-40}	&	{ 93}	&	{ 73}	&	{ CH$_3$OH}	\\
{ 156.300-158.120}	&	{ IRAM 30m}	&	{ 15}	&	{ 50}	&	{ 35-40}	&	{ 93}	&	{ 73}	&	{ CH$_3$OH}	\\
\hline
\end{tabular}
\label{log_obs}
\end{table*}

\section{Results}

\subsection{Spectra}

\begin{figure}[htp]
\centering
\includegraphics[width=\columnwidth]{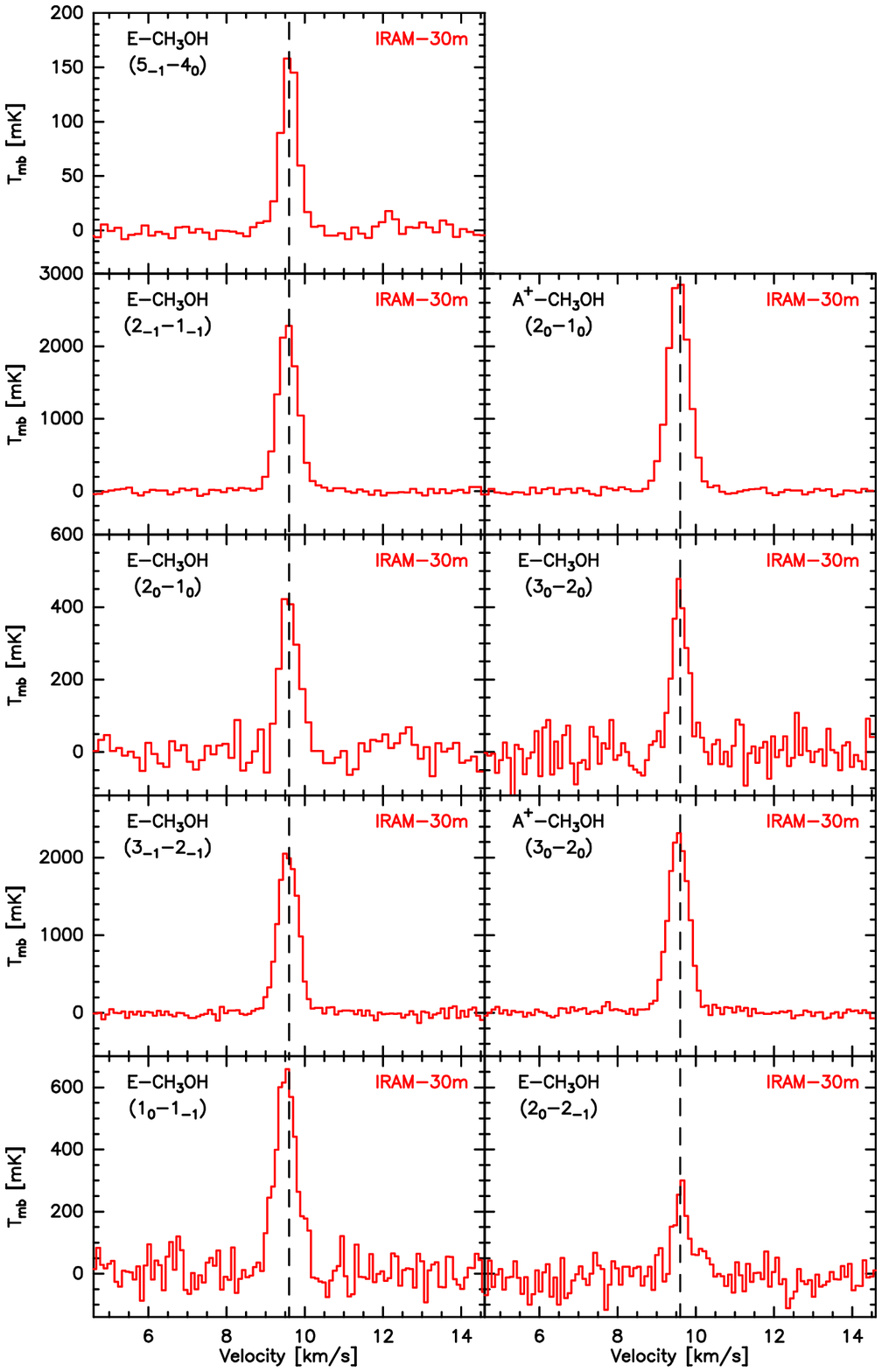}
\caption{Spectrum of the targeted CH$_3$OH transitions towards the B5 methanol hotspot.}
\label{obs_me}
\end{figure}

\begin{figure}[htp]
\centering
\includegraphics[width=\columnwidth]{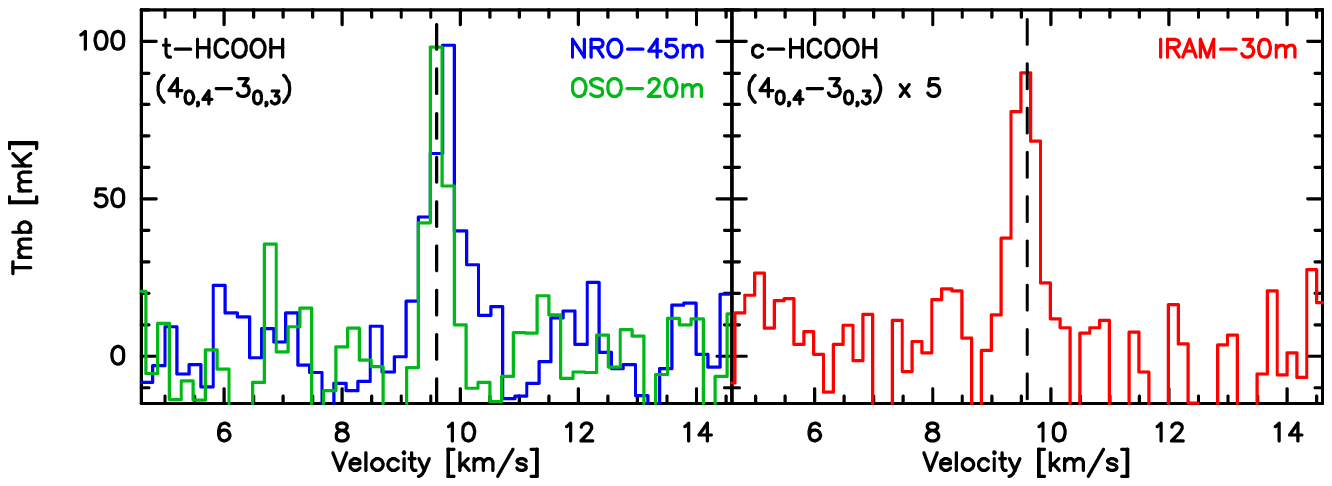}
\caption{Spectra of the targeted HCOOH transitions towards the B5 methanol hotspot.}
\label{obs_fa}
\end{figure}

\begin{figure}[htp]
\centering
\includegraphics[width=45mm]{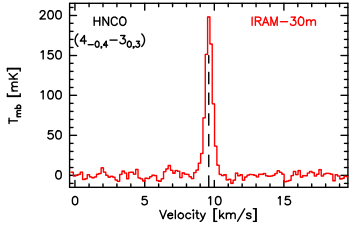}
\caption{Spectrum of the targeted HNCO transition at 89.579 GHz towards the B5 methanol hotspot.}
\label{obs_hnco}
\end{figure}

\begin{figure}[htp]
\centering
\includegraphics[width=\columnwidth]{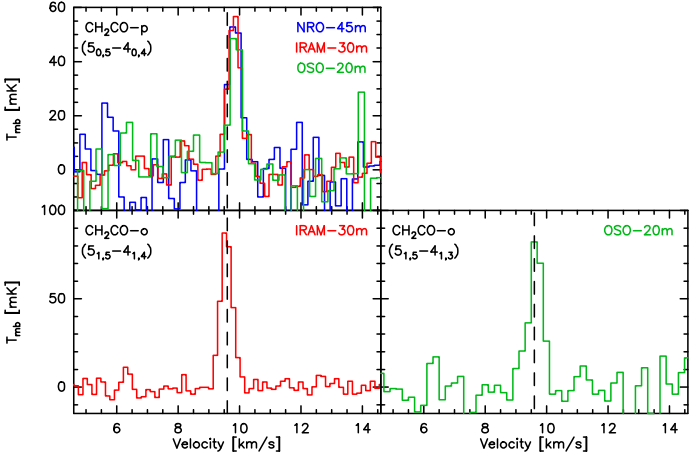}
\caption{Spectra of the targeted CH$_2$CO transitions at $\sim 101$ GHz towards the B5 methanol hotspot.}
\label{obs_ke}
\end{figure}

\begin{figure}[htp]
\centering
\includegraphics[width=\columnwidth]{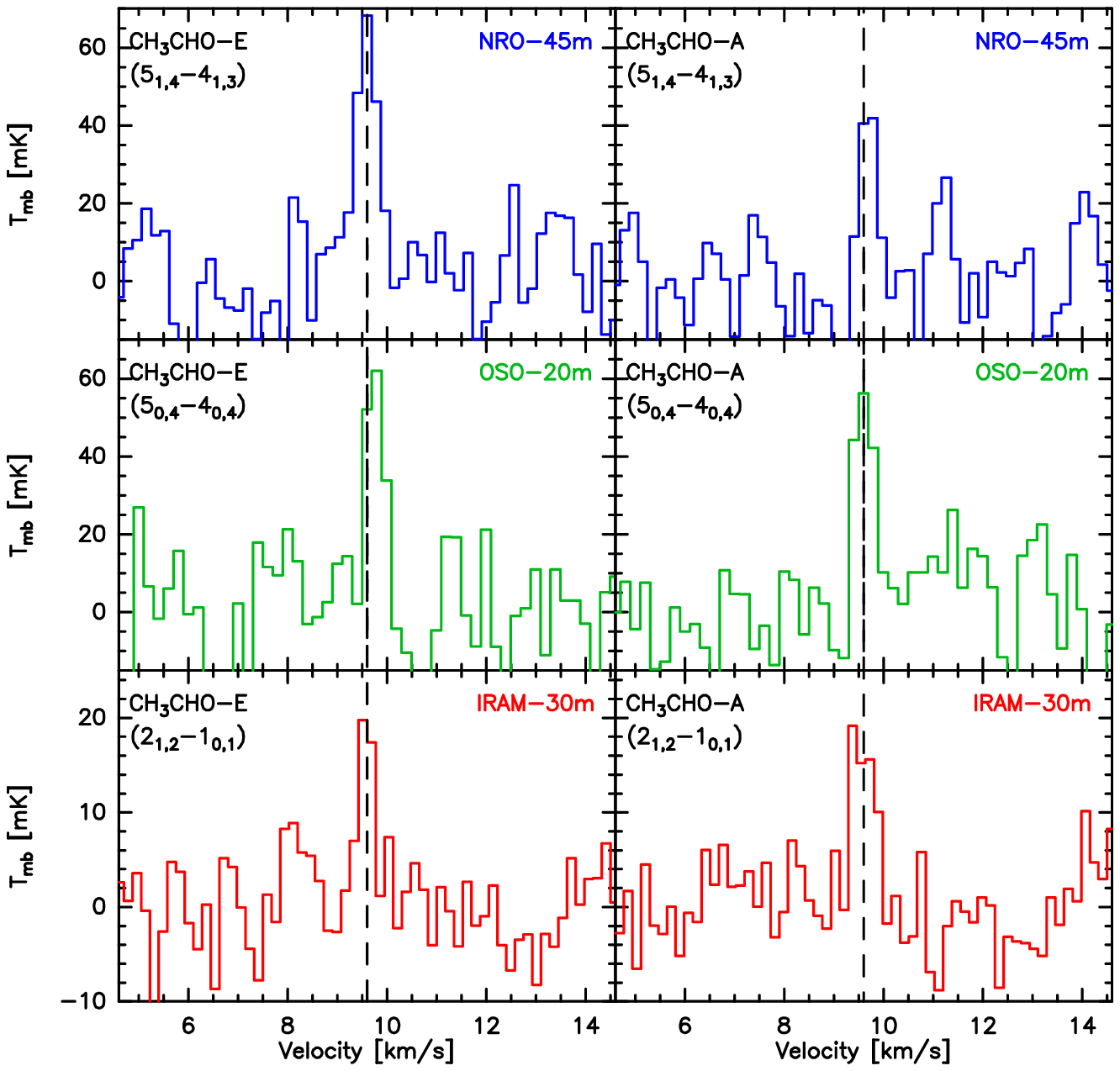}
\caption{Spectra of the targeted CH$_3$CHO transitions towards the B5 methanol hotspot.}
\label{obs_ac}
\end{figure}

\begin{figure}[htp]
\centering
\includegraphics[width=0.6\columnwidth]{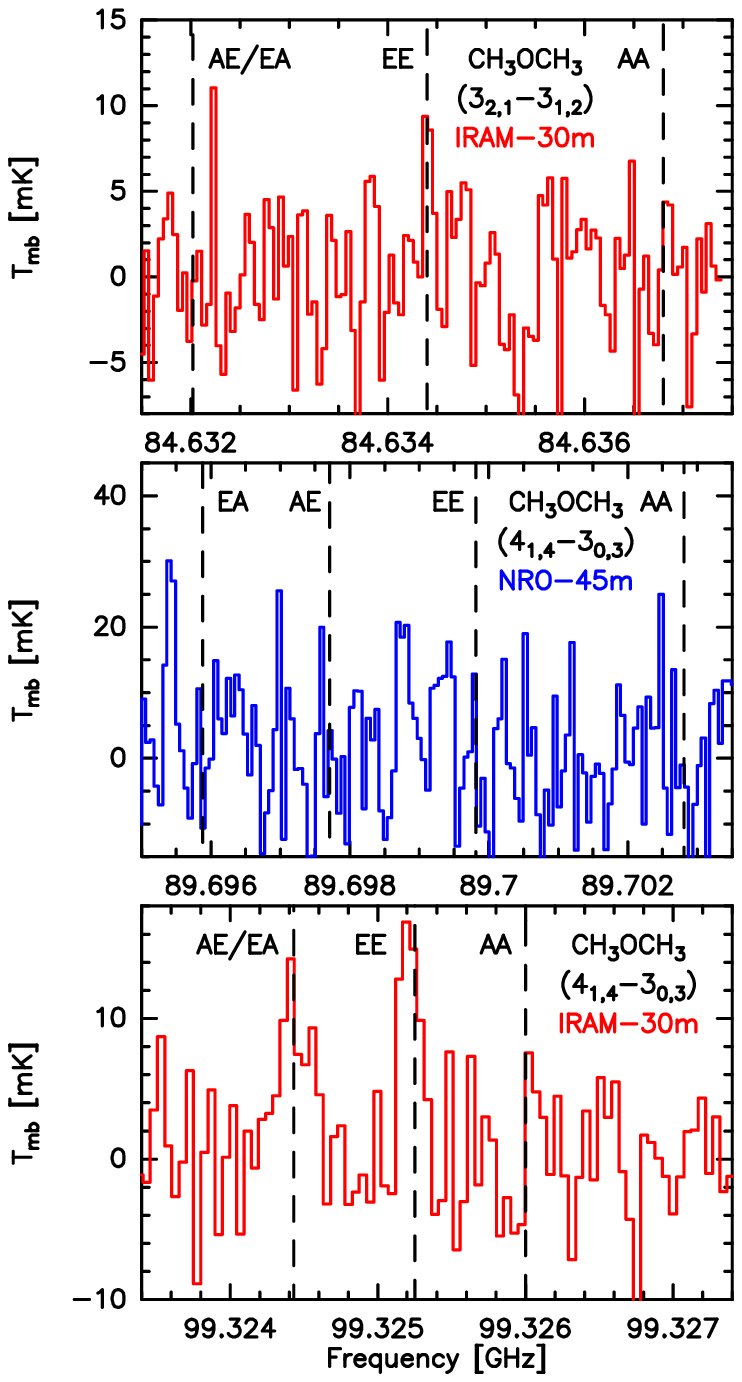}
\caption{Spectra of the targeted CH$_3$OCH$_3$ transitions towards the B5 methanol hotspot.}
\label{obs_dme}
\end{figure}

\begin{figure*}[htp]
\centering
\includegraphics[width=180mm]{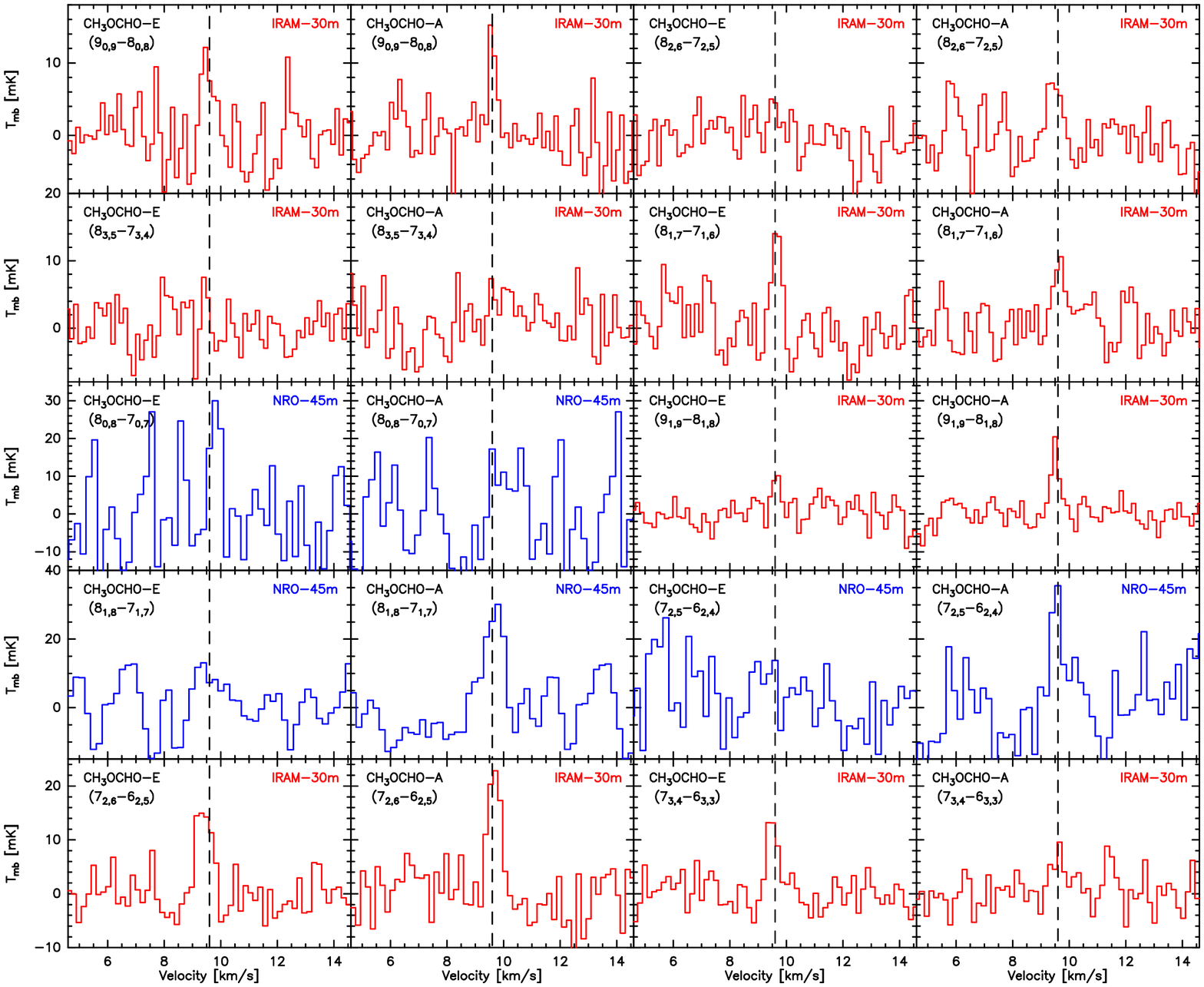}
\caption{Spectra of the targeted CH$_3$OCHO transitions towards the B5 methanol hotspot.}
\label{obs_mf}
\end{figure*}

Figures \ref{obs_me} to \ref{obs_mf} present the spectra obtained with
the NRO 45m (in blue), the IRAM 30m (in red), and the OSO 20m (in
green) telescopes. Table \ref{table_lines} presents the frequencies,
the spectroscopic parameters, and the observed properties of
the targeted transitions. The rms achieved in the observations ranges between
3 and 8 mK per 50-60 kHz spectral bin { for deep observations of
  COMs and between 27 - 40 mK for observations of bright methanol
  transitions}, depending on the observed frequency and the used telescope.  
We report here the detection of several transitions from methanol
(CH$_3$OH), formic acid (HCOOH), ketene (CH$_2$CO), acetaldehyde
(CH$_3$CHO), methyl formate (CH$_3$OCHO), and the tentative detection
of dimethyl ether (CH$_3$OCH$_3$). 

We detected the $4_{0,4} - 3_{0,3}$ transition of HCOOH in its two
conformers {\it trans} and {\it cis} at 89.579 and 87.695 GHz,
respectively (Fig. \ref{obs_fa}). { To our knowledge, this is the second published detection
of the higher energy {\it cis}-HCOOH conformer in space after the
detection by \citet{Cuadrado2016} towards the Orion bar. }
Three transitions from ketene, two in its {\it ortho} substate and one in its
{\it para} substate, have been detected (Fig. \ref{obs_ke}). The {\it para} transition of ketene at
101.036 GHz has been detected with all telescopes. The
integrated main-beam temperatures obtained with the three telescopes
are very similar, the differences in the peak main-beam temperatures
remaining within the uncertainties. This comparison suggests that the
size of the formic acid and ketene emissions is much larger than 40
arcsec, the beam of the OSO 20m at 90 GHz.  In the following, we
therefore assume that the emission size is much larger than the
telescope beams for all the considered species, resulting in a beam
dilution of 1. 

Six acetaldehyde transitions, three for each substate A and E, have
also been detected with upper level energies between 5 and 17 K
(Fig. \ref{obs_ac}). 
CH$_3$OCH$_3$ is detected in the four substates AA, EE, AE, and EA of
the $4_{1,4}-3_{0,3}$ transition whilst the $3_{2,1}-3_{1,2}$
transition at 84  GHz is detected for the substate EE only (Fig. \ref{obs_dme}). The
non-detection of the transitions from the other states is in good
agreement with observations of other dimethyl ether transitions
suggesting that the EE substate usually shows the brightest transition 
\citep[see][]{Bacmann2012}. 
Among the 20 transitions from methyl formate targetted with the IRAM
30m and the NRO 45m telescopes, 13 transitions have been detected at
3$\sigma$ level, 5 for the E substate and 8 in the A substate (Fig. \ref{obs_mf}).
The non-detection of 7 transitions, with similar properties, is 
in good agreement with the high uncertainties of the observed
transitions, detected at a signal-to-noise ratio slightly higher than
3 and showing large uncertainties.  

The Full-Width-at-Half-Maximum (FWHM) of the detected transitions shows a
large variation between 0.3 and 0.7 km/s, especially for the
faint transitions with a low signal-to-noise. Brighter transitions
from methanol or ketene show a FWHM linewidth between 0.5 and 0.6 km/s, as
expected for transitions originated from cold dark clouds. Linewidths
in B5 tend to be slightly larger than the linewidth of 0.4 km/s for the
transitions of COMs detected by \citet{Bacmann2012} towards L1689B and
\citet{JimenezSerra2016} towards L1544.
{ For non-detected transitions, we defined a 3$\sigma$ upper
  limit as $3 \sigma \sqrt{{FWHM} \times \Delta v}$ where $\sigma $ is
  the rms noise, $FWHM$ is the typical linewidth of the
  transition, and $\Delta v$ is the spectral velocity resolution. }

\begin{table*}[htp]
\centering
\caption{List of targeted transitions with their { spectroscopic
    and observed properties. }}
\begin{tabular}{l c c c c c c c c}
\hline
\hline
Molecule	&	Transition	&	Frequency	&	$E_{\textrm{up}}$	&		$A_{\rm{i,j}}$				&	Telescope	&		$\int T_{\textrm{mb}}^*dv$			&	$V{\textrm{peak}}$	&	FWHM	\\
	&		&	(GHz)	&	(K)	&		(s$^{-1}$)				&		&		(mK km/s)			&	(km/s)	&	(km/s)	\\
\hline
E-CH$_3$OH	&	$5_{-1} - 4_{0}$	&	84.52117	&	40.4	&	$	2.0	\times	10^{-6}	$	&	IRAM-30m	&		88.8	$\pm$	17.9	&	9.62	&	0.50	\\
E-CH$_3$OH	&	$2_{-1} - 1_{-1}$	&	96.73936	&	12.5	&	$	2.6	\times	10^{-6}	$	&	IRAM-30m	&		1490	$\pm$	300	&	9.57	&	0.61	\\
A$^+$-CH$_3$OH	&	$2_{0} - 1_{0}$	&	96.74137	&	7.0	&	$	3.4	\times	10^{-6}	$	&	IRAM-30m	&		1950	$\pm$	390	&	9.56	&	0.62	\\
E-CH$_3$OH	&	$2_{0} - 1_{0}$	&	96.74454	&	20.1	&	$	3.4	\times	10^{-6}	$	&	IRAM-30m	&		256	$\pm$	53	&	9.60	&	0.55	\\
E-CH$_3$OH	&	$3_0 - 2_0$	&	145.09375	&	27.1	&	$	1.2	\times	10^{-5}	$	&	IRAM-30m	&		209	$\pm$	44	&	9.59	&	0.47	\\
E-CH$_3$OH	&	$3_{-1} - 2_{-1}$	&	145.09744	&	19.5	&	$	1.1	\times	10^{-5}	$	&	IRAM-30m	&		1250	$\pm$	250	&	9.57	&	0.60	\\
A$^+$-CH$_3$OH	&	$3_0 - 2_0$	&	145.10319	&	13.9	&	$	1.2	\times	10^{-5}	$	&	IRAM-30m	&		1420	$\pm$	284	&	9.56	&	0.61	\\
E-CH$_3$OH	&	$1_0 - 1_{-1}$	&	157.27083	&	15.4	&	$	2.2	\times	10^{-5}	$	&	IRAM-30m	&		410	$\pm$	84	&	9.52	&	0.65	\\
E-CH$_3$OH	&	$2_0 - 2_{-1}$	&	157.27606	&	20.1	&	$	2.2	\times	10^{-5}	$	&	IRAM-30m	&		114	$\pm$	27	&	9.63	&	0.43	\\
\hline
HNCO	&	$4_{0,4}-3_{0,3}$	&	87.92525	&	10.5	&	$	8.5	\times	10^{-6}	$	&	IRAM-30m	&		126	$\pm$	25	&	9.63	&	0.61	\\
\hline
$t$-HCOOH	&	$4_{0,4} - 3_{0,3}$	&	89.57917	&	10.8	&	$	7.2	\times	10^{-6}	$	&	NRO-45m	&		56.8	$\pm$	15.9	&	9.74	&	0.68	\\
	&		&		&		&						&	OSO-20m	&		39.2	$\pm$	6.5	&	9.70	&	0.39	\\
$c$-HCOOH	&	$4_{0,4} - 3_{0,3}$	&	87.69469	&	10.5	&	$	2.5	\times	10^{-5}	$	&	IRAM-30m	&		11.4	$\pm$	3.3	&	9.50	&	0.56	\\
\hline
o-CH$_2$CO	&	$5_{1,5} - 4_{1,4}$	&	100.09451	&	27.5	&	$	1.0	\times	10^{-5}	$	&	IRAM-30m	&		45.8	$\pm$	9.2	&	9.58	&	0.61	\\
p-CH$_2$CO	&	$5_{0,5} - 4_{0,4}$	&	101.03671	&	14.5	&	$	1.1	\times	10^{-5}	$	&	NRO-45m	&		29.5	$\pm$	8.8	&	9.85	&	0.50	\\
	&		&		&		&						&	IRAM-30m	&		31.4	$\pm$	6.4	&	9.83	&	0.52	\\
	&		&		&		&						&	OSO-20m	&		25.6	$\pm$	6.5	&	9.88	&	0.46	\\
o-CH$_2$CO	&	$5_{1,4} - 4_{1,3}$	&	101.98140	&	27.7	&	$	1.1	\times	10^{-5}	$	&	OSO-20m	&		46.8	$\pm$	11.1	&	9.66	&	0.52	\\
\hline
E-CH$_3$CHO	&	$2_{1,2}-1_{0,1}$	&	83.58426	&	5.03	&	$	2.4	\times	10^{-6}	$	&	IRAM-30m	&		8.4	$\pm$	2.1	&	9.57	&	0.36	\\
A-CH$_3$CHO	&	$2_{1,2}-1_{0,1}$	&	84.21976	&	4.96	&	$	2.4	\times	10^{-6}	$	&	IRAM-30m	&		11.4	$\pm$	3.3	&	9.57	&	0.56	\\
E-CH$_3$CHO	&	$5_{0,5}-4_{0,4}$	&	95.94744	&	13.9	&	$	3.0	\times	10^{-5}	$	&	OSO-20m	&		31.4	$\pm$	8.7	&	9.76	&	0.43	\\
A-CH$_3$CHO	&	$5_{0,5}-4_{0,4}$	&	95.96346	&	13.8	&	$	3.0	\times	10^{-5}	$	&	OSO-20m	&		30.4	$\pm$	8.5	&	9.62	&	0.46	\\
E-CH$_3$CHO	&	$5_{1,4}-4_{1,3}$	&	98.86331	&	16.6	&	$	3.0	\times	10^{-5}	$	&	NRO-45m	&		20.0	$\pm$	6.9	&	9.70	&	0.37	\\
A-CH$_3$CHO	&	$5_{1,4}-4_{1,3}$	&	98.90094	&	16.5	&	$	3.0	\times	10^{-5}	$	&	NRO-45m	&		38.1	$\pm$	9.5	&	9.59	&	0.54	\\
\hline
AE/EA-CH$_3$OCH$_3$	&	$3_{2,1}-3_{1,2}$	&	84.63202	&	11.1	&	$	2.2	\times	10^{-6}	$	&	IRAM-30m	&	$<$	3.5			&		&		\\
EE-CH$_3$OCH$_3$	&	$3_{2,1}-3_{1,2}$	&	84.63440	&	11.1	&	$	2.2	\times	10^{-6}	$	&	IRAM-30m	&		4.0	$\pm$	1.4	&	9.56	&	0.34	\\
AA-CH$_3$OCH$_3$	&	$3_{2,1}-3_{1,2}$	&	84.6368	&	11.1	&	$	2.2	\times	10^{-6}	$	&	IRAM-30m	&	$<$	3.5			&		&		\\
EA-CH$_3$OCH$_3$	&	$2_{2,1}-2_{1,2}$	&	89.69588	&	8.4	&	$	1.9	\times	10^{-6}	$	&	NRO-45m	&	$<$	9.4			&		&		\\
AE-CH$_3$OCH$_3$	&	$2_{2,1}-2_{1,2}$	&	89.69771	&	8.4	&	$	1.9	\times	10^{-6}	$	&	NRO-45m	&	$<$	9.4			&		&		\\
EE-CH$_3$OCH$_3$	&	$2_{2,1}-2_{1,2}$	&	89.69981	&	8.4	&	$	1.9	\times	10^{-6}	$	&	NRO-45m	&	$<$	9.4			&		&		\\
AA-CH$_3$OCH$_3$	&	$2_{2,1}-2_{1,2}$	&	89.70281	&	8.4	&	$	1.9	\times	10^{-6}	$	&	NRO-45m	&	$<$	9.4			&		&		\\
AE/EA-CH$_3$OCH$_3$	&	$4_{1,4}-3_{0,3}$	&	99.32443	&	10.2	&	$	4.4	\times	10^{-6}	$	&	IRAM-30m	&		3.9	$\pm$	2.5	&	9.57	&	0.39	\\
EE-CH$_3$OCH$_3$	&	$4_{1,4}-3_{0,3}$	&	99.32525	&	10.2	&	$	4.4	\times	10^{-6}	$	&	IRAM-30m	&		9.2	$\pm$	2.3	&	9.69	&	0.46	\\
AA-CH$_3$OCH$_3$	&	$4_{1,4}-3_{0,3}$	&	99.32600	&	10.2	&	$	4.4	\times	10^{-6}	$	&	IRAM-30m	&		2.8	$\pm$	1.3	&	9.41	&	0.27	\\
\hline
E-CH$_3$OCHO	&	$7_{2,6}-6_{2,5}$	&	84.44917	&	19.0	&	$	8.0	\times	10^{-6}	$	&	IRAM-30m	&		11.4	$\pm$	3.3	&	9.38	&	0.65	\\
A-CH$_3$OCHO	&	$7_{2,6}-6_{2,5}$	&	84.45475	&	19.0	&	$	8.0	\times	10^{-6}	$	&	IRAM-30m	&		13.1	$\pm$	3.5	&	9.70	&	0.47	\\
E-CH$_3$OCHO	&	$7_{3,4}-6_{3,3}$	&	87.14328	&	22.6	&	$	7.7	\times	10^{-6}	$	&	IRAM-30m	&		6.9	$\pm$	2.7	&	9.50	&	0.43	\\
A-CH$_3$OCHO	&	$7_{3,4}-6_{3,3}$	&	87.16129	&	22.6	&	$	7.8	\times	10^{-6}	$	&	IRAM-30m	&		4.7	$\pm$	2.6	&	9.62	&	0.58	\\
E-CH$_3$OCHO	&	$8_{1,8}-7_{1,7}$	&	89.31466	&	20.2	&	$	1.0	\times	10^{-5}	$	&	NRO-45m	&	$<$	5.9			&		&		\\
A-CH$_3$OCHO	&	$8_{1,8}-7_{1,7}$	&	89.31664	&	20.1	&	$	1.0	\times	10^{-5}	$	&	NRO-45m	&		22.5	$\pm$	7.3	&	9.67	&	0.72	\\
E-CH$_3$OCHO	&	$7_{2,5}-6_{2,4}$	&	90.14572	&	19.7	&	$	9.8	\times	10^{-6}	$	&	NRO-45m	&	$<$	9.9			&		&		\\
A-CH$_3$OCHO	&	$7_{2,5}-6_{2,4}$	&	90.15647	&	19.7	&	$	9.8	\times	10^{-6}	$	&	NRO-45m	&		17.3	$\pm$	8.2	&	9.55	&	0.46	\\
E-CH$_3$OCHO	&	$8_{0,8}-7_{0,7}$	&	90.22766	&	20.1	&	$	1.1	\times	10^{-5}	$	&	NRO-45m	&	$<$	9.9			&		&		\\
A-CH$_3$OCHO	&	$8_{0,8}-7_{0,7}$	&	90.22962	&	20.1	&	$	1.1	\times	10^{-5}	$	&	NRO-45m	&	$<$	9.9			&		&		\\
E-CH$_3$OCHO	&	$9_{1,9}-8_{1,8}$	&	100.07861	&	24.9	&	$	1.4	\times	10^{-5}	$	&	IRAM-30m	&		3.4	$\pm$	1.4	&	9.65	&	0.31	\\
A-CH$_3$OCHO	&	$9_{1,9}-8_{1,8}$	&	100.08054	&	24.9	&	$	1.5	\times	10^{-5}	$	&	IRAM-30m	&		6.0	$\pm$	1.7	&	9.50	&	0.30	\\
E-CH$_3$OCHO	&	$8_{3,5}-7_{3,4}$	&	100.29460	&	27.4	&	$	1.3	\times	10^{-5}	$	&	IRAM-30m	&	$<$	2.8			&		&		\\
A-CH$_3$OCHO	&	$8_{3,5}-7_{3,4}$	&	100.30818	&	27.4	&	$	1.3	\times	10^{-5}	$	&	IRAM-30m	&	$<$	3.1			&		&		\\
E-CH$_3$OCHO	&	$8_{1,7}-7_{1,6}$	&	100.48224	&	22.8	&	$	1.4	\times	10^{-5}	$	&	IRAM-30m	&		7.5	$\pm$	1.9	&	9.66	&	0.36	\\
A-CH$_3$OCHO	&	$8_{1,7}-7_{1,6}$	&	100.49068	&	22.8	&	$	1.4	\times	10^{-5}	$	&	IRAM-30m	&		5.4	$\pm$	1.6	&	9.60	&	0.64	\\
E-CH$_3$OCHO	&	$9_{0,9}-8_{0,8}$	&	100.68154	&	24.9	&	$	1.5	\times	10^{-5}	$	&	IRAM-30m	&		6.3	$\pm$	1.7	&	9.54	&	0.70	\\
A-CH$_3$OCHO	&	$9_{0,9}-8_{0,8}$	&	100.68337	&	24.9	&	$	1.5	\times	10^{-5}	$	&	IRAM-30m	&		4.4	$\pm$	1.5	&	9.59	&	0.23	\\
E-CH$_3$OCHO	&	$8_{2,6}-7_{2,5}$	&	103.46657	&	24.6	&	$	1.5	\times	10^{-5}	$	&	IRAM-30m	&	$<$	3.1			&		&		\\
A-CH$_3$OCHO	&	$8_{2,6}-7_{2,5}$	&	103.47866	&	24.6	&	$	1.5	\times	10^{-5}	$	&	IRAM-30m	&		5.6	$\pm$	1.6	&	9.50	&	0.50	\\
\hline
\end{tabular}
\label{table_lines}
\end{table*}

\subsection{Non-LTE analysis}

\begin{figure*}[htp]
\centering
\includegraphics[width=0.3\textwidth]{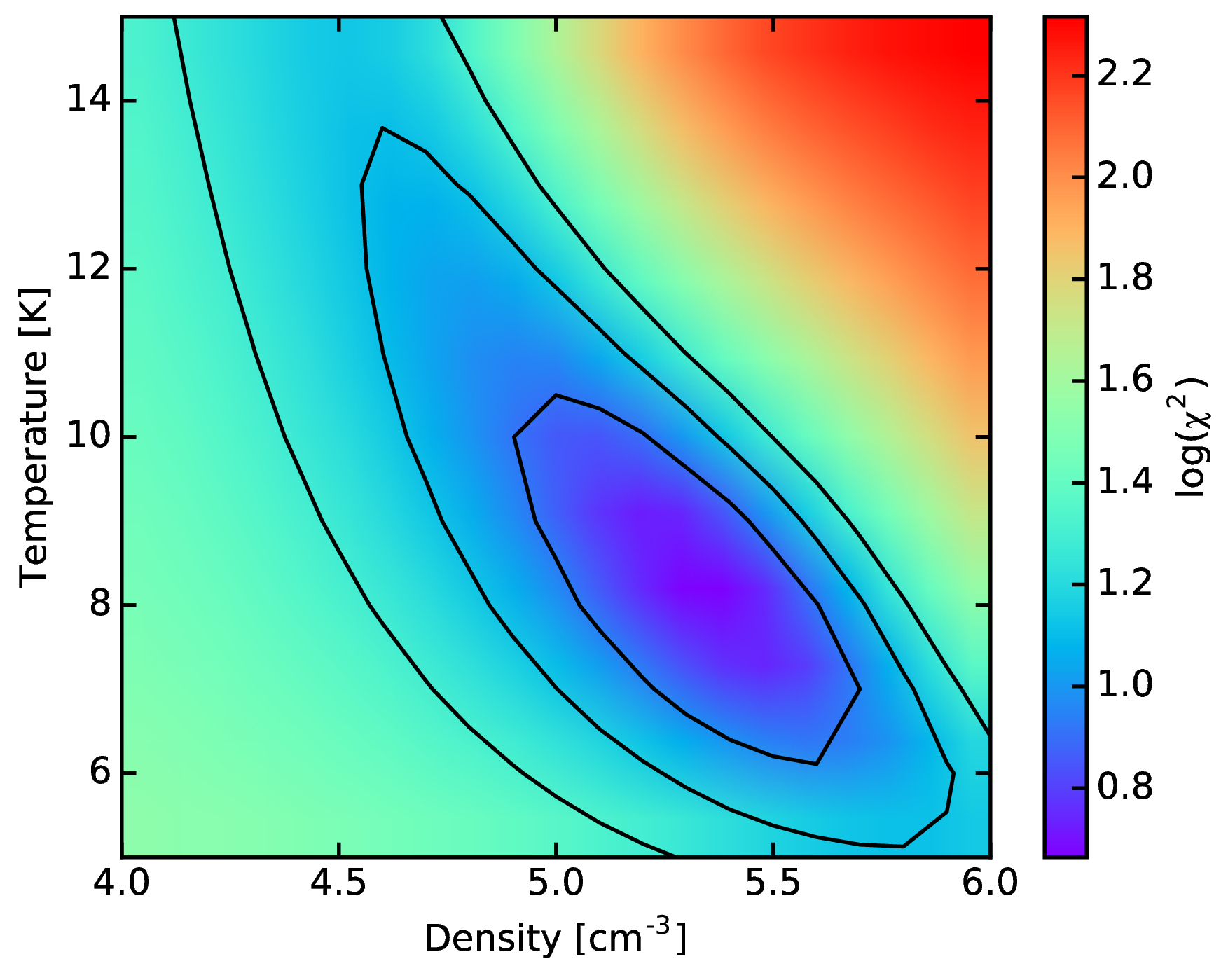}
\includegraphics[width=0.3\textwidth]{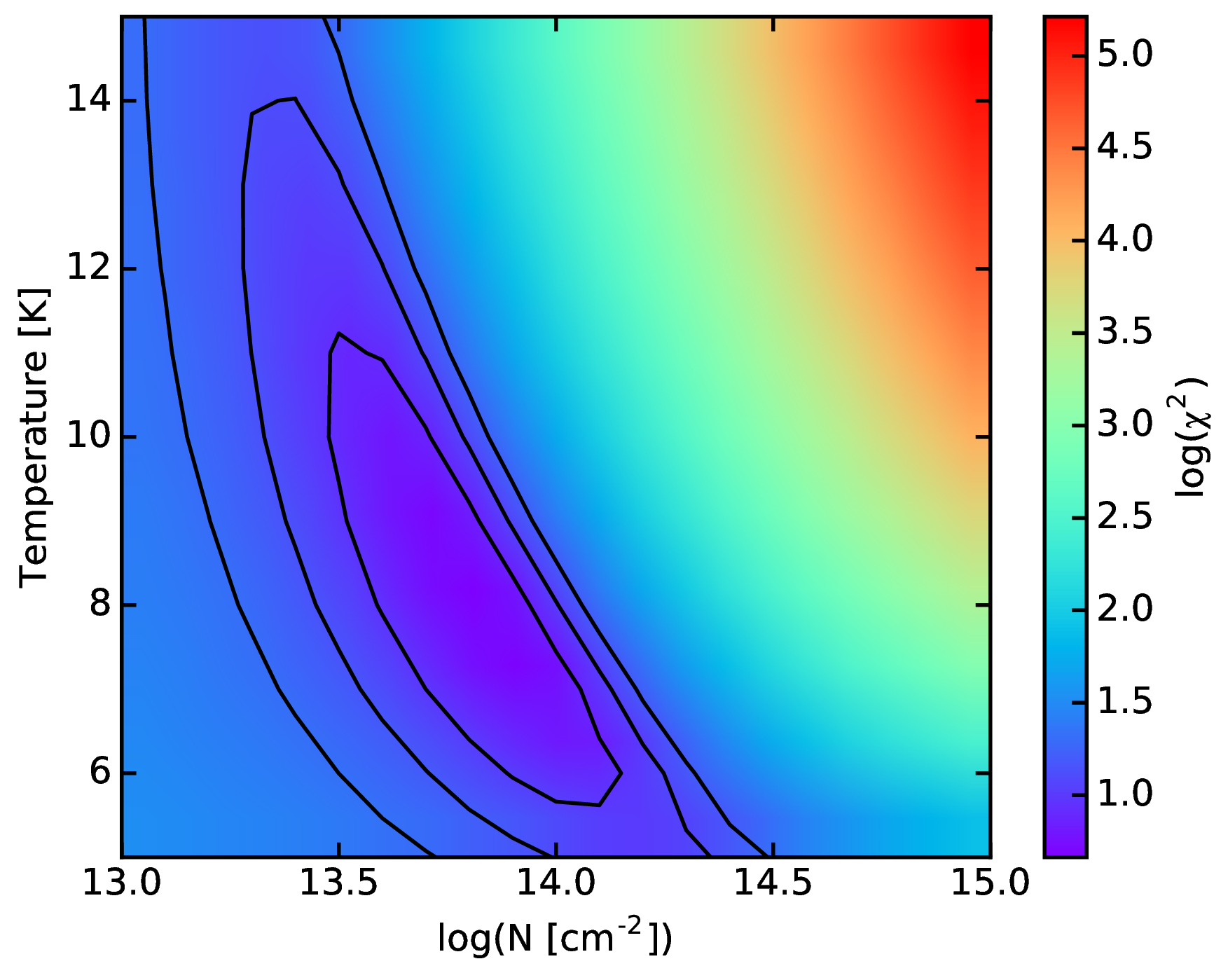}
\includegraphics[width=0.3\textwidth]{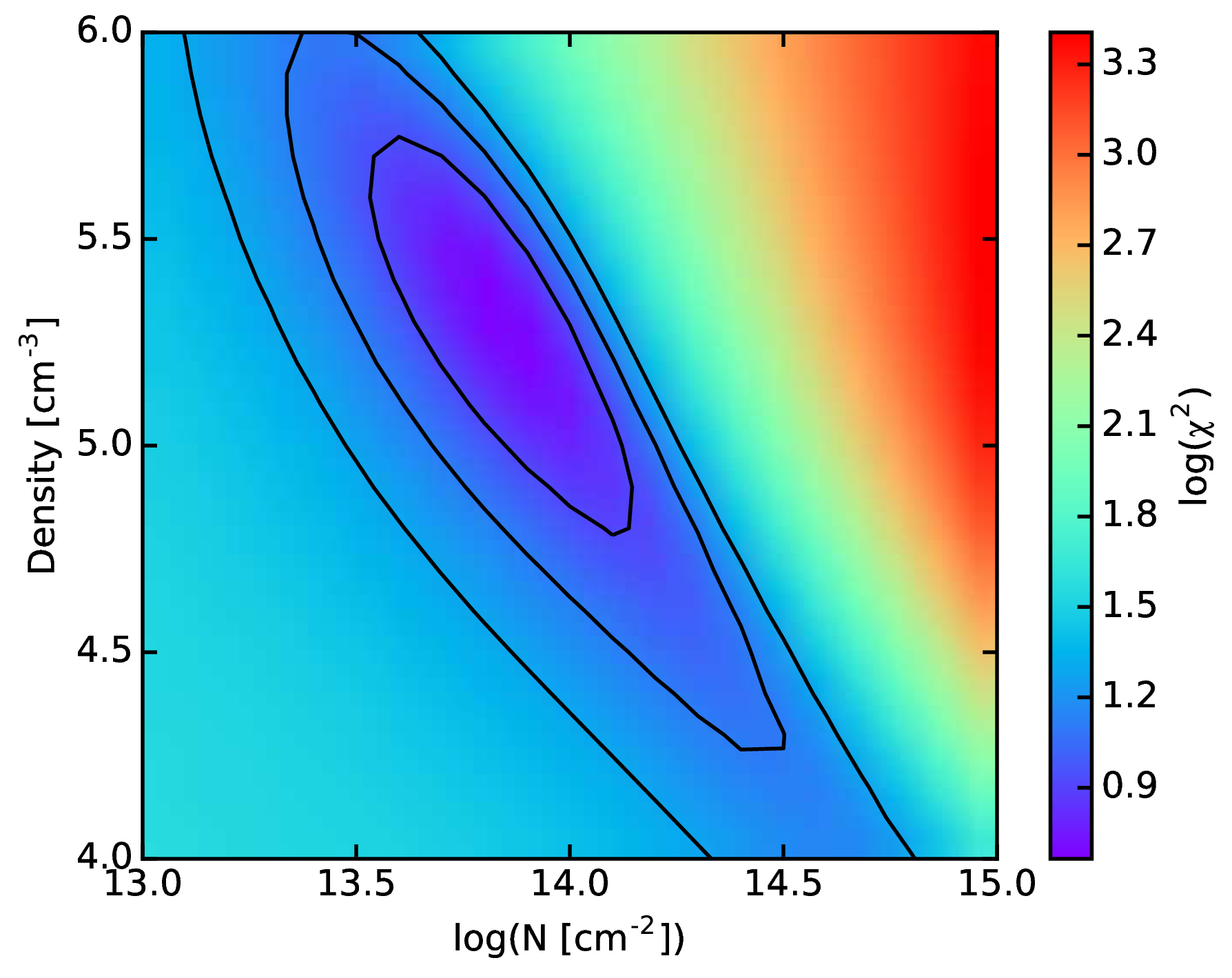}
\caption{$\chi^2$ distribution in the H$_2$ density - kinetic temperature, column
  density - kinetic temperature, and column density - H$_2$ density planes for
  the RADEX fitting of the methanol emission. Black lines show
  $\chi^2$ contours at 1$\sigma$, 2$\sigma$, and 3$\sigma$.}
\label{LVG_ME}
\end{figure*}

Cross-sections for the rotational excitation of the A- and E-types of
methanol and methyl formate by H$_2$ and Helium, respectively, and of
HNCO by H$_2$ have been computed by \citet{Rabli2010},
\citet{Faure2014}, and \citet{Green1986}, allowing us
to perform a non-LTE analysis of the emission of these molecules. 
{ To this mean, we used the non-LTE radiative transfer code RADEX 
assuming an isothermal and homogeneous medium using the large velocity
approximation for a uniform sphere \citep{vanderTak2007}.
The line excitation temperatures $T_{{\rm ex}}$, the opacities $\tau$,
and the integrated brightness temperatures $T_{\rm mb}$ { of
  methanol} have first been computed by varying the column density of
each sub-type $N_{\rm tot}$ between $10^{13}$ and $10^{15}$ cm$^{-2}$
with 20 logarithmic steps, the density of H$_2$ $n_{\rm H2}$ between $10^4$ and
$10^6$ cm$^{-3}$ with 20 logarithmic steps, and 
the kinetic temperature $T_{\rm kin}$ between 5 and 15 K with 10
linear steps.   
Due to the low number of detected A-CH$_3$OH transitions, we assumed a
state E/A abundance ratio of 1 following the observations by
\citet{Bacmann2016} towards a sample of prestellar cores.  
The FWHM linewidth was fixed to 0.5 km s$^{-1}$ following our observed
spectra.
Figure \ref{LVG_ME} presents the reduced $\chi^2$ distribution in the
$T_{\rm kin}$ - $n_{\rm H2}$, $T_{\rm kin}$ - $N_{\rm tot}$, and
$N_{\rm tot}$ - $n_{\rm H2}$ planes. A finer grid was then performed
around the best-fit values with linear steps. 
Methanol observations can be reproduced with a reduced $\chi^2$ of 4.5
for $T_{\rm kin} = 7.5 \pm 1.5$ K, $n_{\rm H2} = 2.25 \pm 1.50 \times
10^5$ cm$^{-3}$ and for a methanol column density of $N$(A-CH$_3$OH) =
$N$(E-CH$_3$OH) = $7.5 \pm 3.0 \times 10^{13}$ cm$^{-2}$. 
The best-fit density is a factor of 2-4 higher than the average
density in the East 286 dense core, the denser of the two nearby cores
\citep{Sadavoy2013}, whilst the kinetic temperature is 2-3 K and 4-5 K
lower than the average dust temperatures of East 286 and East 189,
respectively. 
This finding would suggest that the methanol emission mostly
originates from the center part of the cores, where the physical
conditions are dense and cold. A dedicated analysis of the continuum
emission profile of the two dense cores needs to be carried out in order to
derive their physical structure.  
%
For these physical conditions and this column density, the transitions
with integrated brightness temperatures higher than 1 K km s$^{-1}$
have an opacity $\tau$ of 1-2 showing that some CH$_3$OH transitions
can be optically thick in B5.  The excitation temperatures vary
between negative values for the $5_{-1}-4_0$, E transition at 84.521 GHz
to $\sim 7 - 7.5$ K for the E and A$^+$ transitions at $\sim 96$
GHz. Appendix 1 discusses the impact of physical
conditions on the excitation temperatures in more detail. 
}

{
The narrow range of upper level energies of the methyl formate
transitions (between 19 and 25 K) prevents us from running a full
model grid in which several parameters are varied, since it is
impossible to converge towards one single solution that reproduces the
observations.  
Instead, we used the H$_2$ density and the temperature that best
  reproduce the methanol emission and we only varied the methyl
  formate column density, first between $10^{11}$ and $10^{15}$
  cm$^{-2}$ and then on a finer grid between $5 \times 10^{11}$ and $5 \times 10^{12}$
  cm$^{-2}$. We did not assume any state E/A abundance
  ratio and the A- and E- states were treated as two distinct but
  coexistent species. 
Methyl formate observations can be reproduced with a reduced $\chi^2$
of $\sim 1$ for $N$(E-CH$_3$OCHO) = $2.1 \pm 1.5 \times 10^{12}$
cm$^{-2}$ and $N$(A-CH$_3$OCHO) = $2.2 \pm 1.3 \times 10^{12}$
cm$^{-2}$. 
For this density, all transitions are optically thin ($\tau << 1$) and
show an excitation temperature $T_{\rm ex}$ of $6-7$ K, close to the
kinetic temperature. 
The HNCO column density that best reproduces the emission of the
transition at 87.92525 GHz is $2.0 \pm 0.6 \times 10^{11}$ cm$^{-2}$. 
The results of the RADEX analysis are summarised in Table
\ref{RADEX_summary}.
}

\begin{table*}[htp]
\centering
\caption{Results of the RADEX analysis of the methanol and methyl
  formate emissions.}
\begin{tabular}{l c c c c c}
\hline
\hline																	
Molecule	&		$N_{\textrm{tot}}$	&	$N_{\textrm{tot}}$/$N$(CH$_3$OH)	&	$N_{\textrm{tot}}$/$N$(H$_2$)	&		$T_{\textrm{kin}}$			&	$n_{\rm H2}$	\\
	&		(cm$^{-2}$)	&		&		&		(K)			&	(cm$^{-3}$)	\\
\hline												
A-CH$_3$OH	&		7.5(+13)	$\pm$	3.0(+13)	&		&	2.3(-8)	&		7.5	$\pm$	1.5	&	2.25(+5) $\pm$ 1.50(+5)	\\
E-CH$_3$OH	&		7.5(+13)	$\pm$	3.0(+13)	&		&	2.3(-8)	&		7.5	$\pm$	1.5	&	2.25(+5) $\pm$ 1.50(+5)	\\
CH$_3$OH	&		1.5(+14)	$\pm$	4.2(+13)	&		&	4.5(-8)	&		7.5	$\pm$	1.5	&	2.25(+5) $\pm$ 1.50(+5)	\\
\hline											
A-CH$_3$OCHO	&		2.2(+12)	$\pm$	1.3(+12)	&	1.5(-2)	&	6.7(-10)	&		7.5$^a$			&	2.25(+5)$^a$	\\
E-CH$_3$OCHO	&		2.1(+12)	$\pm$	1.5(+12)	&	1.4(-2)	&	6.4(-10)	&		7.5$^a$			&	2.25(+5)$^a$	\\
CH$_3$OCHO	&		4.3(+12)	$\pm$	2.0(+12)	&	2.9(-2)	&	1.3(-9)	&					&		\\
\hline																	
HNCO	&		2.0(+12)	$\pm$	6.0(+11)	&	1.3(-2)	&	6.1(-10)	&		7.5$^a$			&	2.25(+5)$^a$	\\
\hline
\end{tabular}
\tablebib{
$^a$: The physical conditions are fixed following the results of the
methanol analysis.}
\label{RADEX_summary}
\end{table*}

\subsection{LTE analysis}

\begin{figure}
\centering
\includegraphics[width=80mm]{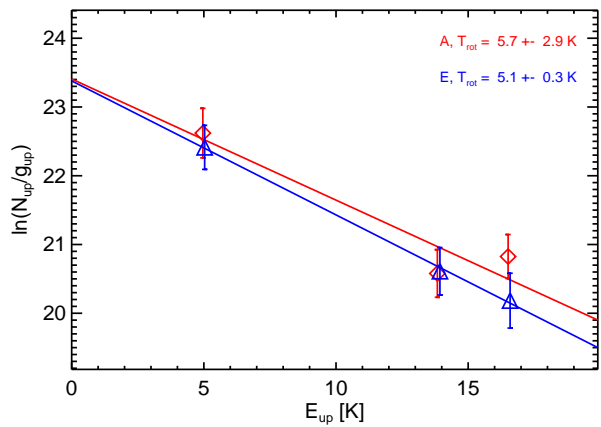}
\caption{Rotational diagram of acetaldehyde CH$_3$CHO. A- and E-states
  are depicted by red diamonds and blue triangles, respectively. }  
\label{rotdiag}
\end{figure}

{ 
The column density of the observed molecules whose collisional
coefficients are not known has been obtained through the Local Thermodynamic
Equilibrium (LTE) approximation.  }
For the kinetic temperatures ($T_{\textrm{kin}} \sim 8$ K) and densities
($n_{\textrm{H}} \sim 10^5$ cm$^{-3}$) expected towards the methanol
hotspot of Barnard 5, the different substates have to be treated
separately. 
{ The detected acetaldehyde transitions have upper level energies 
between 5 and 16 K, allowing us to perform a Population Diagram
analysis. 
The column densities and the rotational temperatures of the two
substates are varied to fit the observational data using the routine
presented in \citet{Taquet2015} following the Population Diagram
method proposed by \citet{Goldsmith1999}.  }
Table \ref{resultsLTE} lists the column densities derived at the
methanol hotspot of Barnard 5 { with the LTE analysis} and Figure
\ref{rotdiag} presents the Rotational Diagram of acetaldehyde. 
{ The rotational temperatures of the two acetaldehyde states are about 5-6
K. This confirms that the observed acetaldehyde transitions are not fully thermalised
for such dark cloud conditions. 
}

The transitions of ketene and dimethyl ether have
similar upper level energies, { preventing us to perform a
  Rotational Diagram analysis in which the rotational temperature is
  considered as a free parameter.} We therefore only varied the total
column density in order to obtain the best fit to the observations assuming two
excitation temperatures: 7.5 K, which corresponds to the kinetic 
temperature in the methanol hotspot { derived with the RADEX analysis of
the methanol emission}, and a lower excitation temperature of 5.0 K
close to the { methanol and methyl formate excitation temperatures, and
of the acetaldehyde rotational temperature}, as should be expected for
molecular cloud conditions where levels are usually sub-thermally
excited.
{ The four different substates of di-methyl ether were treated
  separately. The EA- and AE- transitions have the same frequency at
  99.32443 GHz, we therefore considered that the contribution of the
  two states to the detected transition scales with the degeneracy of
  their upper energy levels. }

{ As expected, the A/E abundance ratio is close to 1 for methyl formate
and acetaldehyde.  The ortho/para ratio of ketene is equal to $3.3 \pm
1.2$, which is consistent with the statistical ratio of 3 and the
value of 3.3 and 3.5 found in L1689B and TMC1 
\citep{Bacmann2012, Ohishi1991}. }

We also derived the upper limits of other undetected COMs showing at
least one bright (i.e., low $E_{\textrm{up}}$, high $A_{\rm{ij}}$)
transition lying in the observed frequency bands. In particular,
glycol aldehyde HCOCH$_2$OH, an isomer of methyl formate, and ethanol,
the ethyl counterpart of methanol, show abundance upper limits that
are about 10 times lower  than the species detected in this work.

\begin{table*}[htp]
\centering
\caption{Column densities and abundances of targetted molecules
  assuming LTE conditions.}
\begin{tabular}{l c c c c c c}
\hline
\hline																	
Fixed $T_{\rm rot}$	&		\multicolumn{3}{c}{$T_{\textrm{rot}} = 8$ K}	&		\multicolumn{3}{c}{$T_{\textrm{rot}} = 5$ K}	\\
\hline															
Molecule	&		$N_{\textrm{tot}}$	&	$N_{\textrm{tot}}$/$N$(CH$_3$OH)	&	$N_{\textrm{tot}}$/$N$(H$_2$)	&		$N_{\textrm{tot}}$	&	$N_{\textrm{tot}}$/$N$(CH$_3$OH)	&	$N_{\textrm{tot}}$/$N$(H$_2$)	\\
	&		(cm$^{-2}$)	&		&		&		(cm$^{-2}$)	&		&		\\
\hline															\\
{\it trans}-HCOOH	&		2.6(+12)	$\pm$	6.6(+11)	&	1.7(-2)	&	7.9(-10)	&		7.5(+12)	$\pm$	1.9(+12)	&	5.0(-2)	&	2.3(-9)	\\
{\it cis}-HCOOH	&		1.7(+11)	$\pm$	4.6(+10)	&	1.1(-3)	&	5.2(-11)	&		4.7(+11)	$\pm$	1.3(+11)	&	3.1(-3)	&	1.4(-10)	\\
HCOOH	&		2.8(+12)	$\pm$	6.6(+11)	&	1.8(-2)	&	8.4(-10)	&		8.0(+12)	$\pm$	1.9(+12)	&	5.3(-2)	&	2.4(-9)	\\
\hline																			
o-CH$_2$CO	&		2.1(+12)	$\pm$	4.2(+11)	&	1.4(-2)	&	6.4(-10)	&		4.8(+12)	$\pm$	9.6(+11)	&	3.2(-2)	&	1.5(-9)	\\
p-CH$_2$CO	&		6.2(+11)	$\pm$	1.8(+11)	&	4.1(-3)	&	1.9(-10)	&		1.5(+12)	$\pm$	4.3(+11)	&	1.0(-2)	&	4.5(-10)	\\
CH$_2$CO	&		2.7(+12)	$\pm$	4.6(+11)	&	1.8(-2)	&	8.2(-10)	&		6.3(+12)	$\pm$	1.1(+12)	&	4.2(-2)	&	1.9(-9)	\\
\hline																			
AA-CH$_3$OCH$_3$	&		3.1(+11)	$\pm$	1.5(+11)	&	2.1(-3)	&	9.4(-11)	&		4.2(+11)	$\pm$	2.0(+11)	&	2.8(-3)	&	1.3(-10)	\\
EE-CH$_3$OCH$_3$	&		1.3(+12)	$\pm$	4.4(+11)	&	8.7(-3)	&	3.9(-10)	&		1.7(+12)	$\pm$	6.1(+11)	&	1.1(-2)	&	5.2(-10)	\\
AE-CH$_3$OCH$_3$	&		2.2(+11)	$\pm$	2.3(+11)	&	1.5(-3)	&	6.7(-11)	&		2.9(+11)	$\pm$	1.6(+11)	&	1.9(-3)	&	8.8(-11)	\\
EA-CH$_3$OCH$_3$	&		2.2(+11)	$\pm$	2.3(+11)	&	1.5(-3)	&	6.7(-11)	&		2.9(+11)	$\pm$	1.6(+11)	&	1.9(-3)	&	8.8(-11)	\\
CH$_3$OCH$_3$	&		2.1(+12)	$\pm$	5.7(+11)	&	1.4(-2)	&	6.2(-10)	&		2.7(+12)	$\pm$	6.8(+11)	&	1.8(-2)	&	8.2(-10)	\\
\hline																			
CH$_3$CCH	&	$<$ 	7.3(+11)			&	4.9(-3)	&	2.2(-10)	&	$<$ 	7.3(+12)			&	4.9(-2)	&	2.2(-9)	\\
HCOCH$_2$OH	&	$<$ 	2.7(+11)			&	1.8(-3)	&	8.3(-11)	&	$<$ 	7.3(+11)			&	4.9(-3)	&	2.2(-10)	\\
aGg-(CH$_2$OH)$_2$	&	$<$ 	4.9(+11)			&	3.2(-3)	&	1.5(-10)	&	$<$ 	2.3(+12)			&	1.5(-2)	&	7.0(-10)	\\
gGg-(CH$_2$OH)$_2$	&	$<$ 	1.2(+12)			&	8.2(-3)	&	3.7(-10)	&	$<$ 	3.5(+12)			&	2.4(-2)	&	1.1(-9)	\\
C$_2$H$_5$OH	&	$<$ 	4.1(+11)			&	2.8(-3)	&	1.3(-10)	&	$<$ 	1.0(+12)			&	6.7(-3)	&	3.1(-10)	\\
NH$_2$CHO	&	$<$ 	2.0(+10)			&	1.3(-4)	&	6.1(-12)	&	$<$ 	5.4(+10)			&	3.6(-4)	&	1.6E(-11)	\\
\hline																			
Rotational diagram	&		\multicolumn{6}{c}{Free $T_{\textrm{rot}}$}										\\
\hline																	
	&		$N_{\textrm{tot}}$	&	$N_{\textrm{tot}}$/$N$(CH$_3$OH)	&	$N_{\textrm{tot}}$/$N$(H$_2$)	&		$T_{\textrm{rot}}$			&		&		\\
	&		(cm$^{-2}$)	&		&		&		(K)			&		&		\\
\hline																			
A-CH$_3$CHO	&		2.7(+12)	$\pm$	5.5(+11)	&	1.8(-2)	&	8.2(-10)	&		5.7	$\pm$	2.9	&		&		\\
E-CH$_3$CHO	&		2.5(+12)	$\pm$	4.6(+11)	&	1.7(-2)	&	7.6(-10)	&		5.1	$\pm$	0.3	&		&		\\
CH$_3$CHO	&		5.2(+12)	$\pm$	7.2(+11)	&	3.5(-2)	&	1.6(-9)	&					&		&		\\
\hline
\end{tabular}
\tablebib{
$^a$: The abundance ratio relative to CH$_3$OH is computed with the
CH$_3$OH column density derived from the RADEX analysis
($N$(CH$_3$OH) = $1.2 \times 10^{14}$ cm$^{-2}$; see text). 
$^b$: The H$_2$ column density is equal to $3.3 \times
  10^{21}$ cm$^{-2}$ \citep{Wirstrom2014}.}
\label{resultsLTE}
\end{table*}

\section{Discussion}

\subsection{Comparison with other sources}

\begin{figure*}
\centering
\includegraphics[width=0.8\textwidth]{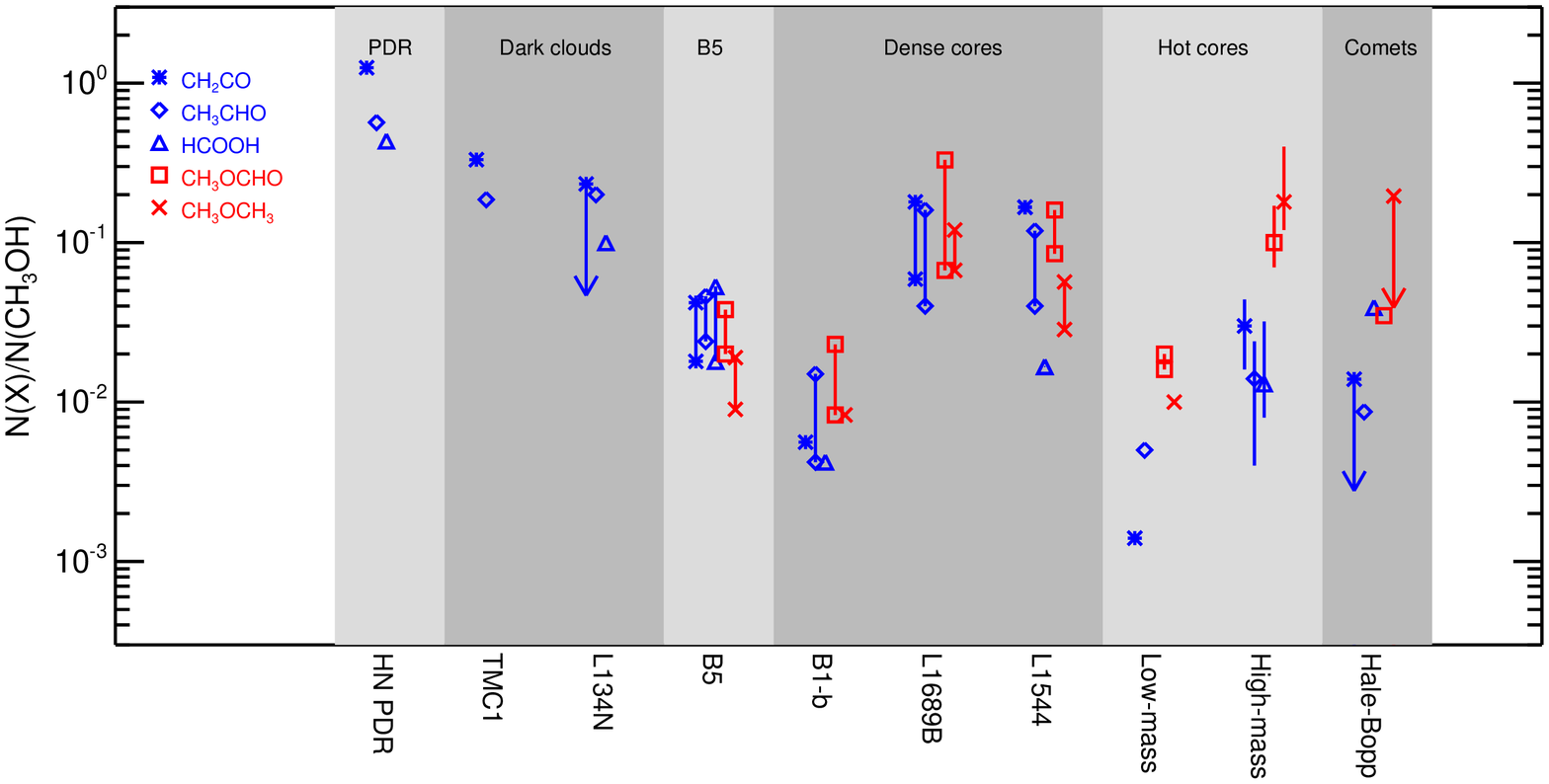}
\caption{Abundances of ``lukewarm'' (blue) and ``warm'' (red) COMs with
  respect to methanol from PDR regions to comets. Abundances in the
  Horsehead Nebula Photo-Dissociated Region (HN PDR) are taken from
  \citet{Guzman2014}. Abundances in the TMC1 and L134N dark clouds are
  from \citet{Ohishi1992} and \citet{Gratier2016}. Abundances in the
  B1-b, L1689B, and L1544 dense cores are taken from
  \citet{Oberg2010}, \citet{Cernicharo2012}, \citet{Bacmann2012, Bacmann2016}, and
  \citet{Vastel2014, Bizzocchi2014, JimenezSerra2016}. Abundances in low-mass protostars are those
  derived in NGC1333-IRAS2A by \citet{Taquet2015} for CH$_2$CO,
  CH$_3$OCHO, and CH$_3$OCH$_3$ and in IRAS 16293-2422-A by
  \citet{Jorgensen2016} for CH$_3$CHO, and CH$_3$OCHO. Abundances in
  high-mass hot cores are the mean values and the standard deviations
of abundances derived from about 20 high-mass hot cores and compiled
in \citet{Taquet2015}. Abundances in the comet Hale-Bopp are those
derived by \citet{BockeleeMorvan2000}.}
\label{Xch3oh_coms}
\end{figure*}

The methanol hotspot in B5 is located just between two dense cores and
therefore represents a stage between the dark cloud and dense core phases.
As a consequence, abundances of COMs with respect to methanol, {
  their likely parent species}, derived
towards B5 listed in Table \ref{resultsLTE} are compared in Figure
\ref{Xch3oh_coms} with observations towards the dark clouds TMC1 and
L134N and the dense cores B1-b, L1689B, and L1544 
\citep{Ohishi1992, Oberg2010, Cernicharo2012, Bacmann2012,
  Bizzocchi2014, Vastel2014, Bacmann2016, Gratier2016, JimenezSerra2016}.
Abundances derived at the ``methanol hotspot'' of B5 seem to be in
good agreement with observations towards other cold dense cores. 
In B5, B1-b, and L1689B, all the detected COMs have similar abundances
within each source, at least considering the ranges of possible
abundances depending on their assumed excitation temperature. However,
COMs in B1-b are found to be less abundant with respect to CH$_3$OH
than in L1689B by $\sim$ one order of magnitude.  
L1544 is the only source showing significant abundance differences
between COMs: { ketene is ten times more abundant than formic acid
  and two to four times more abundant than acetaldehyde.}

Figure \ref{Xch3oh_coms} also shows abundance ratios of COMs derived
in other types of sources. 
The Horsehead Nebula Photon-Dominated Region (HN PDR) observed by
\citet{Guzman2014} is located in the external layer of the Horsehead
nebula and undergoes efficient UV photochemistry. It therefore
represents the translucent phase prior the formation of dark clouds. 
Hot cores represent the early stages of star formation
when the central source is still embedded in an thick protostellar
envelope. Abundances in low-mass protostars are the values derived
towards NGC1333-IRAS2A and IRAS 16293-2422 through sub-mm interferometric
observations by \citet{Taquet2015} and \citet{Jorgensen2016}, whilst abundances
in high-mass protostars are averaged values derived from the compilation by
\citet{Taquet2015}.
Abundances in the comet Hale-Bopp by \citet{BockeleeMorvan2000} are
representative of abundances at the end of the protoplanetary disk
phase.
Although many other phases are involved in the star formation process,
such a comparison can give us a first idea of the abundance evolution of
different types of COMs with the evolutionary stage of star formation.    
The five COMs studied in this work can be distinguished into two
categories according to their rotational temperature around massive
hot cores \citep[see][]{Isokoski2013}:
the ``lukewarm'' COMs ketene, acetaldehyde, and formic acid 
showing low rotational temperatures between 20 and 80 K; and
the ``warm'' COMs methyl formate and di-methyl ether showing 
rotational temperatures usually higher than 100 K.  

The three ``lukewarm'' COMs are found to be as abundant as methanol in the
Horsehead PDR but they show abundances of 10 - 30 \% in dark clouds
and abundances mostly lower than 10 \% in cold dense cores. 
Around low- and high-mass hot cores, the ``warm'' COMs methyl formate
and di-methyl ether are more abundant than the three ``lukewarm'' COMs. 
The abundance of CH$_2$CO and CH$_3$CHO derived in the hot corino of
NGC1333-IRAS2A and IRAS 16293-2422 by \citet{Taquet2015} and
\citet{Jorgensen2016} are lower than 0.5 \% whereas abundances of
CH$_3$OCHO and CH$_3$OCH$_3$ are higher than 1 \%. 
Abundances of ``lukewarm'' COMs around massive hot cores derived
from single-dish observations and assuming similar source sizes for all
molecules are about 1 \% with respect to methanol, and could be even
lower if a larger source size is assumed as expected from their low
excitation temperature, whilst CH$_3$OCHO and CH$_3$OCH$_3$ abundances
are higher than 10 \%. 

\subsection{Ice evaporation processes in dark clouds}

Methanol cannot form in the gas phase \citep{Luca2002,
  Garrod2006} whilst water is the main component of interstellar 
ices \citep[see][]{Oberg2011}.
The detection of cold methanol and water with high abundances of $4.5
\times 10^{-8}$, derived with our RADEX analysis, and $\sim 2 \times
10^{-8}$ w.r.t. H$_2$ \citep{Wirstrom2014} respectively, in a dark 
cloud region with no associated infrared emission is therefore
attributed to cold grain surface chemistry followed by non-thermal
evaporation processes.   
Several processes have been proposed to trigger evaporation of icy
material in the gas phase of dark clouds: 
1) photoevaporation induced by the external interstellar and cosmic
ray induced field of UV photons;
2) chemical desorption upon surface formation of the molecule;
3) ice sputtering by cosmic-rays.
{ A viable desorption mechanism must also  account for the 
spatial chemical differentiations and anti-correlations observed for
the different molecules detected in Barnard 5. Processes that are not
able to explain the observed spatial distributions can therefore
be discounted. }

UV photolysis of methanol ice has been experimentally studied by
\citet{Bertin2016} and \citet{CruzDiaz2016}. They showed that UV
photolysis of methanol ice mostly results in methanol
photodissociation. \citet{Bertin2016} derived a low photoevaporation
rate of $\sim 10^{-5}$ molecules per incident photon for a pure
methanol ice, two orders of magnitude lower than the rates commonly
used in astrochemical models. The photoevaporation is found to be an
indirect process in which the radicals produced by photodissocation
recombine together to release the produced methanol molecule into the gas
phase through chemical desorption. UV photolysis of a more realistic
methanol-CO ice mixture decreases the rate to $< 3 \times 10^{-6}$
molecules per incident photon. 
The location of the methanol hotspot between two dense cores that can
act as shields against external UV irradiation, in combination with
its low efficiency, suggests that UV photoevaporation is not the
dominant mechanism responsible for its high gas-phase methanol
abundance. 

The so-called chemical desorption, desorption of a product due to the
energy release of the exothermic surface reaction, has been
experimentally quantified by \citet{Minissale2016} for a dozen of
reactions for an amorphous water ice substrate, representative of ices
observed in dense clouds. 
Although these authors have only been able to derive an upper limit
of 8 \% for the desorption efficiency of methanol due to the reaction
between CH$_3$O and H, their theoretical estimate results in a chemical
desorption efficiency of $\sim 2$ \%. 
A high abundance of gaseous methanol between the two dense cores
East189 and East286 could therefore reflect an efficient methanol
formation at the surface of interstellar ices triggering chemical
desorption of methanol into the gas phase. 

The sputtering efficiency induced by the impact of energetic ions on
cold water ice has been studied experimentally by various authors
\citep[see][]{Brown1984}. Recently, \citet{Dartois2015} measured the
yield of water ice sputtering by swift heavy MeV ions and derived the
sputtering rate of water ice for different cosmic ray ionisation
rates, which was found to vary between 4 and 30 molecules cm$^{-2}$
s$^{-1}$. Assuming that the cosmic-ray induced FUV field is about
$10^4$ photons cm$^{-2}$ s$^{-1}$ \citep{Shen2004}, UV photodesorption would contribute
equally to the cosmic ray sputtering with a photodesorption probability of
10$^{-3}$. Assuming that methanol has a similar sputtering efficiency
than water, this result suggests that ice sputtering by cosmic rays is a much more
efficient process than UV photodesorption, by a factor of 100, for releasing
the ice content in the gas phase. The abundance of gas phase methanol
induced by cosmic ray ice sputtering would therefore only
depend on the total methanol abundance in ices, controlling the total
evaporation rate, and the density of the medium, governing the
accretion rate. As for chemical desorption, the bright methanol emission seen between
two dense cores, and not at the edge of the cores more illuminated by
external UV photons, suggests that methanol ice would form efficiently
in UV-shielded and colder regions. 
{ This result is consistent with the observations of methanol
  towards the L1544 prestellar core by \citet{Bizzocchi2014} and
  \citet{Spezzano2016}. The methanol emission in L1544 peaks at the
  northern part of the core where the interstellar radiation field is
  likely weaker. }

As mentioned by \citet{Wirstrom2014}, other processes, such as
collisions between small gas clumps triggering the heating of
interstellar grains due to grain-grain collisions, might also be at
work. 
{ A detailed analysis of the CH$_3$OH maps taken with the IRAM 30m
  will help us to clarify this issue.}

\subsection{Detection of {\it cis}-HCOOH}

{\it cis}-HCOOH has been detected in Barnard 5 at a
relative abundance to {\it trans}-HCOOH of about 6 \%, { but only
  from the detection of one single transition at a signal-to-noise of 5}. 
The detected $4_{0,4}-3_{0,3}$ transition is the {\it cis}-HCOOH line that is expected to be
strongest in the 3 mm band under dark cloud conditions. Assuming a
population distribution thermalised at $\sim 8 - 10$ K, { only one transition}, the $5_{0,5}-4_{0,4}$
  line at 109.47 GHz, should be detectable at similar noise levels, and
  while our observations do not cover this frequency, the detection
  can thus be relatively easily confirmed or refuted. 
HCOOH has been shown to form from hydrogenation of HO-CO radicals in
interstellar ices at low temperatures \citep{Goumans2008,
  Ioppolo2011}. The {\it cis}-HCOOH isomer lies 1365 cm$^{-1}$ higher
in energy than {\it trans}-HCOOH \citep{Lattanzi2008}, and is
therefore expected to be about 800 times less abundant around 300 K,
and even rarer at the low temperatures of dark clouds. However, a
solid state formation mechanism does not necessarily produce an
equilibrium isomeric distribution, since the hydrogenation outcome
also depends on how the HO-CO complex binds to the surface and its orientation
against the incoming H atom. In addition, the association of OH and CO on a
grain surface  forms HO-CO in a {\it trans} configuration, which can only
isomerise to the slightly higher energy {\it cis}-HOCO by overcoming an
energy barrier \citep{Goumans2008}. Furthermore, { the relative
  orientation of the OH group in the cis/trans isomers of HOCO and
  HCOOH} give that  
\begin{equation} cis-\ce{HOCO} + \ce{H} \rightarrow
trans-\ce{HCOOH} \end{equation}
\begin{equation} trans-\ce{HOCO} + \ce{H} \rightarrow
cis-\ce{HCOOH}  \end{equation}
Thus, it seems possible that a non-negligible fraction of the formic
acid formed on grains should be in the {\it cis}-configuration. 

\citet{Cuadrado2016} recently detected several transitions from {\it
  cis}-HCOOH towards the Orion Bar photodissociation region and
derived a higher {\it cis}-to-{\it trans} abundance ratio of 36
\%. This higher ratio could be due to a
photoswitching mechanism proposed by these authors in which the
absorption of a UV photon by the {\it trans} HCOOH conformer
radiatively excites the molecule to overcome the {\it trans}-to-{\it
  cis} interconversion barrier. As discussed previously, the methanol
hotspot of B5 is likely shielded from strong sources of UV irradiation,
unlike the Orion Bar, this photoswitching process is therefore likely
inefficient in this dark cloud region. 

\subsection{Implication for the formation and the evolution of COMs}

The different trends for the abundance ratio evolution with the
evolutionary stage of star formation are consistent with the
rotational temperatures derived in high-mass hot cores
\citep[see][]{Isokoski2013}. 
The ``lukewarm'' COMs acetaldehyde, ketene, and formic acid should indeed be
mostly formed in cold regions either through gas phase chemistry and/or
at the surface of interstellar grains. 
The difference of abundances between PDRs and dark clouds suggests
that either the three ``lukewarm'' COMs are more efficiently formed with
respect to methanol in transluscent regions than in dark clouds or
that methanol is more efficiently destroyed through UV
photodissociation than ketene, acetaldehyde, and formic acid. 
To our knowledge, the UV photodissociation cross sections of the three
later species are not known and are usually assumed to be
equal to those of methanol.
As mentionned earlier, the detection of {\it cis}-HCOOH suggests that
formic acid would be mostly formed in the inner part of interstellar
ices together with CO$_2$ because both molecules can be formed from the 
reaction between CO+OH through the HO-CO complex \citep{Goumans2008,
  Ioppolo2011} and can then be released into the gas phase via
chemical desorption.  
On the other hand, important gas phase formation pathways have been
proposed for ketene and acetaldehyde.
Cold ketene is mostly formed from the electronic recombination of
CH$_3$CO$^+$ produced by ion-neutral chemistry whilst cold
acetaldehyde would be mostly produced through the neutral-neutral
reaction between C$_2$H$_5$ and O and via the electronic recombination
of CH$_3$CHOH$^+$ produced by ion-neutral reactions such as
CH$_4$+H$_2$CO$^+$ \citep{Occhiogrosso2014}.
As mentioned in the introduction, surface formation routes producing
these two species have also been proposed either through atomic carbon
addition of CO followed by hydrogenation \citep{Charnley1997} or
through radical-radical recombination triggered by the UV photolysis
of the main ice components during the warm-up phase
\citep{Garrod2008}. 
The simultaneous detection in high abundances of CH$_3$CHO and HCOOH
along with H$_2$O and CH$_3$OH is strongly suggestive of a surface origin.
However, the former surface reaction scheme has not been experimentally
confirmed yet whilst the latter is known to be ineffective under
laboratory conditions \citep{Oberg2009}.  

As mentioned in the Introduction, a small fraction of methyl formate
and di-methy ether could be formed in cold conditions, either through
neutral-neutral and ion-neutral reactions in the gas or through
radical recombination induced by the CO hydrogenation in interstellar
ices. 
Cold surface formation of COMs from radical-radical recombination
induced by addition and abstraction reactions of CO, H$_2$CO, and
CH$_3$OH has been experimentally demonstrated by \citet{Fedoseev2015} and
\citet{Chuang2016}. However, these experiments tend to produce more 
glycolaldehyde HCOCH$_2$OH than its isomer methyl formate CH$_3$OCHO,
especially when methanol ice is used as initial substrate, because
abstraction reactions of methanol only occur on the methyl group
favouring the production of CH$_2$OH with respect to CH$_3$O. 
However, glycolaldehyde was not detected towards B5 and we derived a
[HCOCH$_2$OH]/[CH$_3$OCHO] abundance ratio lower than 6 \%. Assuming that
the evaporation efficiency is similar for the two isomers, the low
abundance ratio suggests that radical-radical recombination on ices,
followed by evaporation, is not the dominant mechanism for the
formation of gas phase COMs in dark clouds.
Gas phase chemistry triggered by the evaporation of methanol seems to
be efficient enough to produce methyl formate and di-methyl ether in
similar quantities with abundances of a few percents with respect to methanol
\citep{Balucani2015, JimenezSerra2016}. It should be noted that these
results highly depend on the choice of poorly constrained rate
coefficients for some key gas phase reactions.

The higher abundances of these two COMs in protostellar cores
with respect to the ``lukewarm'' species suggests either that the bulk of
their molecular material is formed afterwards under warmer conditions or
that the ``lukewarm'' COMs are destroyed under warmer conditions. 
High angular resolution observations, allowing us to spatially resolve
the hot cores around high- and low-mass sources, are therefore
needed to follow the evolution of these complex species with the
evolutionary stage of star formation.

\section{Conclusions}

This work presents the detection of several Complex Organic Molecules (COMs)
towards the ``methanol hotspot'' located between two dense cores of
the Barnard 5 Molecular Cloud. Through the use of LTE and non-LTE
methods, we have been able to derive the abundances of the targeted
COMs with respect to methanol, their likely parent molecule. 
We summarise here the main conclusions of this work: \\
1) The efficient non-thermal evaporation of methanol and water is
accompanied by the detection of various COMs in high abundances,
between $\sim 1$ and $\sim 10$ \% with respect to methanol. \\
2) Formic acid has been detected in its two
conformers {\it trans} and {\it cis}. The {\it cis}/{\it trans}
abundance ratio of 6 \% confirms that formic acid would be mostly formed at
the surface of interstellar grains and then released into the gas
phase through non-thermal evaporation processes { since the
  photoswitching mechanism proposed by \citet{Cuadrado2016} can be ruled
  out in this region}. \\
3) { The non-LTE RADEX analysis of the methanol emission allowed us to
constrain precisely the physical conditions associated with the
observed emission. The methanol emission would originate from a dense
($n_{\rm H2} \sim 2 \times 10^5$ cm$^{-3}$) and cold ($T_{\rm kin} = 7-8$ K) gas,
consistent with the center part of dense cores, compatible with the
findings by \citet{Bacmann2016} in a sample of prestellar cores.  } \\ 
4) Our targeted COMs can be defined as ``lukewarm'' or ``warm'' species,
according to their rotational temperatures in a sample of massive
hot cores \citep[see][]{Isokoski2013}. 
Comparison with other observations confirms that ``lukewarm''  and ``warm'' 
COMs form and show similar abundances in low-density cold gas, just as
in dense cores -- unlike in protostellar cores where the “warm” COMs
tend to be more abundant than the ``lukewarm''  species. 
The evolution of abundances from dark clouds to protostellar cores
suggests either that ``warm'' COMs are indeed mostly formed in
protostellar environments and/or that ``lukewarm'' COMs are efficiently
destroyed in warm conditions. \\
5) The low [glycol aldehyde]/[methyl formate] abundance ratio of $< 6$ \% suggests
that surface chemistry is not the dominant mechanism for the formation
of methyl formate, and possibly also of di-methyl ether.

\begin{acknowledgements}

The authors are grateful to the IRAM staff for successfully carrying
out the IRAM 30m observations and to the NRO staff for their excellent
support. V. T. thanks S. Sadavoy for sharing the properties of the
Barnard 5 dense cores and N. Sakai for her help in preparing the NRO
observations. 
Astrochemistry in Leiden is supported by the European Union A-ERC grant 291141
CHEMPLAN.
E.S.W. acknowledges generous financial support from the Swedish
National Space Board. 
S.B.C. is supported by NASA's Emerging Worlds and the Goddard Center
for Astrobiology.

  \end{acknowledgements}

 \newpage

\appendix

\section{Excitation of methanol and methyl formate levels}

\begin{figure}
\centering
\includegraphics[width=\columnwidth]{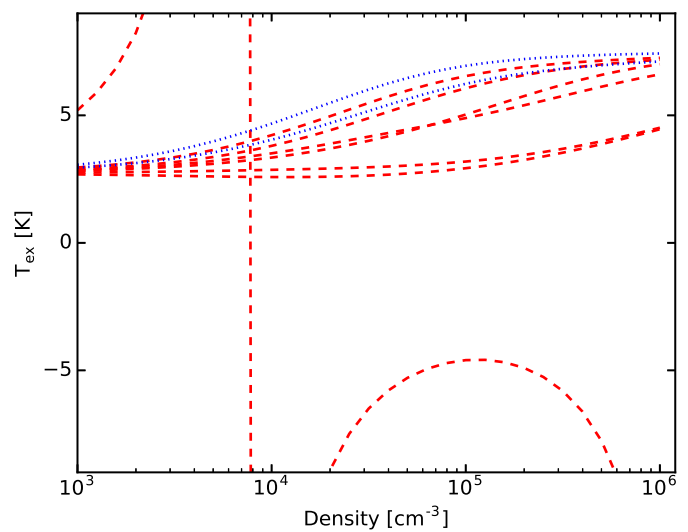}
\caption{Excitation temperatures of the methanol transitions as
  function of the H$_2$ density for $T_{\rm kin} = 7.5$ K. The blue dashed
  curve shows the A-state transition whilst the red solid curves show
  the E-state transitions. The transition showing an inversion
  population is the one at 84.521 GHz.}
\label{Tex_ME}
\end{figure}

\begin{figure}
\centering
\includegraphics[width=\columnwidth]{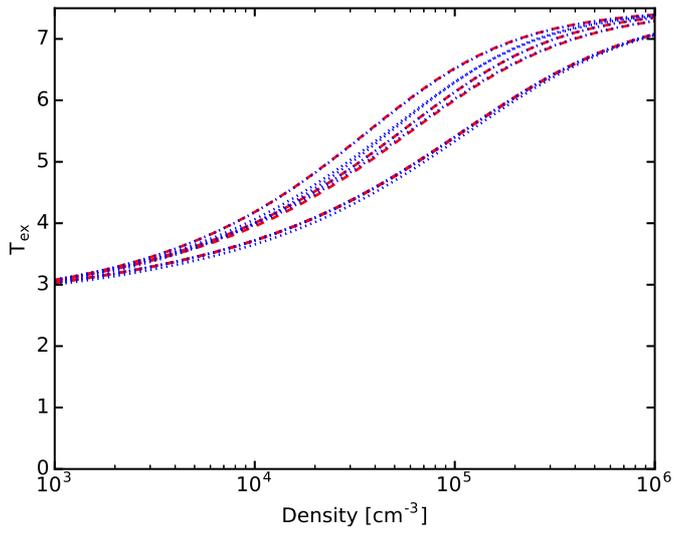}
\caption{Excitation temperatures of the methyl formate transitions as
  function of the H$_2$ density for $T_{\rm kin} = 7.5$ K. The blue dotted
  curves show the A-state transitions whilst the red dashed curves show
  the E-state transitions.}
\label{Tex_MF}
\end{figure}

The excitation temperatures computed with RADEX for the observed methanol and
methyl formate transitions as function of the H$_2$ density are shown in Figures
\ref{Tex_ME} and \ref{Tex_MF} respectively. The column density is the
best-fit value listed in Table \ref{RADEX_summary} and $T_{\rm kin}$
is set to 7.5 K. 
All the methyl formate transitions follow the same trend, their excitation
temperatures increase with density from $\sim 3$ K at $10^3$ cm$^{-3}$
to reach the kinetic temperature at densities higher than $10^6$
cm$^{-3}$ showing that most methyl formate transitions in the 3mm band
should be thermalised for densities higher than $10^6$ cm$^{-3}$. 
{ For methanol, the situation is more complex.
The E-CH$_3$OH transition at 84.521 GHz shows a
negative excitation temperature for densities higher than $10^4$
cm$^{-3}$, showing that this inversion population of the 
$5_{-1} - 4_{0}$ doublet is a robust phenomenon. This transition
is known to be a strong maser since its detection towards the massive
protostar DR21(OH) by \citet{Batrla1988}, explaining its bright
emission in our data. 
Other CH$_3$OH transitions show positive excitation temperatures but
do not follow the same trend with the density. The two $J_{0} -
J_{-1}$, E- transitions at 157 GHz show lower ($\sim 3$ K) excitation
temperatures than other transitions. Their excitation temperatures
remain lower than 5 K even at high densities whilst the excitation
temperature of other transitions is close to the kinetic temperature. 
As already shown in many previous publications, the thermalisation
conditions of the CH$_3$OH transitions strongly depend on their
critical densities, which can highly vary between different upper
level quantum numbers. As an example, the detected $J_{0} - J_{-1}$
($J = 1, 2$), E- transitions have  critical densities higher than $10^6$ cm$^{-3}$
whilst the $2_{K} - 1_{K}$ ($K = -1, 0$), E- transitions show critical densities of
$2-3 \times 10^4$ cm$^{-3}$ at 10 K, explaining their different
excitation temperatures. 
}

\end{document}